# THE STUDY ON PERCEPTUAL TRAINING OF CHINESE MANDARIN TONES FOR MONOLINGUAL SPEAKERS OF ENGLISH USING ADAPTIVE COMPUTER BASED TRAINING SOFTWARE

**Written by**

**Wang Yuke**

**McGill University**

**June 2023**

**Title: The Study of Perceptual Training of Chinese Mandarin Tones for Monolingual Speakers of English Using Adaptive Computer Based Training Software**
**Name: Wang Yuke**

# ABSTRACT

With the rapid development of technology and information science, human-computer interaction and individualized learning have been recognized in modern language training programs. Many second language learners struggled to handle Chinese Mandarin tones in adulthood. It’s an innovative research approach to help American Chinese learners improve their capabilities of perceiving Chinese Mandarin tones through a human-computer interaction based training model. The thesis aimed to enhance cognitive abilities to perceive four Chinese tones for American monolingual speakers of English by utilizing a self-designed software, and analyze the results by comparing training outcomes between a human computer interaction based training model and a traditional training model.

The study selected 10 participants from across 18 to 24 in University of Michigan, distributed them randomly into two groups, and provided them with different training programs separately. The human computer interaction based training software used synthesized materials that are easier for the human brain to perceive and traditional training program utilized same amounts of regular study materials. The results demonstrated that human computer interaction training program was more effective in helping learners to build up the more native-like categorical perception of Chinese Mandarin tones. The fourth tone increased the most, the third tone was most difficult to identify while the second tone was hardest to differentiate with other tones. The study testified the effectiveness of enhancing cognitive capabilities of perceiving Chinese Mandarin tones through training program employing infant-directed speech, visual cues and adaptive listening of materials with variations of difficulties.

The study explored a new technique of phonetic tone training, which may have a positive impact on second language learning and tone training.

# Contents

# Chapter 1 Introduction

## 1.1 Background

We all know that infants are faster learners. A two-year-old infant could perceive the language tones within a short period of time and easily improve their ability to identify or differentiate speech sounds and hence enhance their tone-perceiving capability in a shocking quickly way. In fact, learning occurs among children when they respond to a range of specific stimuli or triggering of information. When they are exposed to a specific external or internal stimulus, their brains start sequencing available information that eventually fosters in meaningful learning. A child's perceptual system is an amazing tool to enhance his or her ability to respond to any given environmental impetus. In other words, perceptual learning is a great way to bring long-lasting and positive changes in the life of a child. Every child also has a definite pattern of learning and usually, it connects perceptual thinking systems. In other words, a child always possesses a particular perceptual thinking pattern that results in productive learning.

Although lots of studies about perceptual training and effectiveness of infant-directed speech for second language phonetic learning have been developed, few articles in major journals address the topic in relation to a computer based perceptual training session with the characteristics of infant-directed speech.

It’s a universally acknowledge truth that Mandarin is difficult to learn. Many individuals including the people who teach it as professionals seem to assume it’s difficult for monolingual speakers of English to comprehend, especially the pitch patterns of the tones. In fact, Mandarin is a tonal language, in which tones are used to distinguish words, and consist of variations of the fundamental frequency (F0), amplitude, and voice quality attached to each syllable at the suprasegmentally level (Zhang 2011). Because tones are crucial to lexical contrast and are phonemic in the same way as vowels and consonants (Heinzen2014), variation in pitch alone can completely change the meaning of a word. In Mandarin, it differentiates four tones in terms of high level (ā), high rising (á), dipping (ǎ), and high falling (à), indicating misuse of tones in Mandarin will result in confusion of understanding of the sentences and ineffectiveness of the communication (Zhang 2010). For example, mā in high level tone means mother however mǎ in dipping tone means horse; mǎi in dipping tone means purchase but mài in high falling tone means sell. Thus, understanding the tones and having capabilities to use correct tons in different scenarios are the key factors to improve Mandarin for monolingual speakers of English.

## 1.2 Organization of the Thesis

The goal of the thesis is to enhance cognitive abilities to perceive four Chinese tones for American monolingual speakers of English by utilizing a self-designed software and analyze the results by comparing training outcomes between a human computer interaction based training model and a traditional training model. In order to achieve the purpose, six chapters in terms of introduction, literature review, methods, results, findings, and conclusion were employed to give empirical illustrations. In introduction chapter, background and organization of the thesis were illustrated. In literature review chapter, thesis demonstrated research backgrounds and crucial concepts including perceptual learning and infant-directed speech. In methods chapter, thesis presented detailed description about experiment design and operation process. In findings and discussion chapter, thesis analyzed, audited and evaluated the data and implemented outcome measures to determine the tone training performances. In conclusion chapter, major findings, implications, and limitations were provided and summarized to give a better illustration to the readers. In the end, references, appendices, and acknowledgments were listed to provide a complete and detailed illustration of the experiments.

# Chapter 2 Literature Review

## 2.1 Research Background

Numerous studies examining tone perception by speakers of various native language background show that speakers of non-tonal languages perform less accurately than speakers of tonal languages on tone discrimination or identification tasks (Hao 2012). Similarly, Wayland and Guion (2004) showed that perceptual discrimination of Thai lexical tone contrasts is challenging for non-tonal (American English) speakers, and is resistant to training, whereas Mandarin Chinese speakers improved after training. In addition, Wayland and Guion (2004) showed that American English learners of Thai outperform naive listeners, even though their performance remains significantly lower than native Thai listeners. It is well-established that native, non-native speakers differ from native speakers in their perception/production of Mandarin lexical tones (Best.C 1999) with the former frequently confusing high rising tone and dipping tone, but empirical studies focusing on the processing and acquisition of tones by non-native learners are the minority as pointed out by some researchers (Wang 2012). The late acquisition and perceptual confusion of high rising tone and dipping tone has been frequently reported for Mandarin-speaking children and adults as well as non-native learners (Wang 2012). Cross-linguistically, listeners whose native language is not a tonal language have been shown to process tones in a nonlinguistic manner, or at least not in terms of linguistic categories. Kuhl et al (2011) showed that American English listeners process tones mainly in the right hemisphere, whereas native speakers use their left hemisphere more, supporting the linguistic–nonlinguistic processing difference. Francis (2008) reported that while Mandarin listeners process native tones in a categorical manner, French listeners do not, even though they are able to perceive acoustic differences between them. French is a non-stress language which does not exploit F0 variations at the lexical level. As a result, French listeners display a low sensitivity to syllable-level prosodic variations involving F0 contours. In other words, they are less well equipped to perceive F0 distinctions at the word level and may experience difficulties in discriminating, identifying, and acquiring Mandarin or Thai tonal distinctions. In contrast, previous linguistic experience with tones appears to facilitate perception of unfamiliar tones in a different language (Wayland and Guion 2004).

## 2.2 Randomized Controlled Experiment

In 1885, Charles Sander Peirce and Joseph Jastrow first introduced the concept of Randomized Controlled trial, which is known as RCT into the field of psychology in the paper "on a small difference in sensation". Later, randomized controlled trial was brought into various fields in terms of education, agriculture, and medicine. In 1935, Ronald A. Fisher published his book "Design of Experiments" where he illustrated detailed examples on "how to design experiments systematically from a statistical point of view" (Conniffe, 1990, p. 87). In this thesis, author utilized RCT as basic scientific experiment method to design an experiment to analyze various features including effectiveness, consistency, and percent of the correctness of both an identification task and a discrimination task for participants who are monolingual speakers of English. After testing the accessibility of 10 participants, they were given numbers from 1 to 10 for each individual. Through the technique of tossing the coin, 5 heads were selected into the experiment group 1, and the rest of 5 participants were included into experiment group 2. The experiment group 1 used an adaptive computer-based training software to train their perception capability on identifying and discriminating Chinese Mandarin tones, while the experiment group 2 used self-designed ordinary training materials including videos, lectures, textbooks, and notes. By comparing the outcomes of pretest and post-test, readers could have a clear understanding of how different training programs functioned on four Chinese Mandarin tones.

## 2.3 Perceptual Learning

Perceptual learning in humans was once assumed to be a phenomenon restricted to the early stages of human development or attributable to changes in high-level cognitive processes. In the case of development, a great deal of neural tuning and reorganization takes place during early childhood, and many experiments have shown that perceptual experience during that time can play a large role in permanently shaping the properties of neural mechanisms(Kuhl 2010). It was traditionally assumed that after that critical period of perceptual development had passed, neural mechanisms at the earliest stages of information processing were no longer plastic and thus could not be modified through experience with the world. In the case of perceptual learning in adults, it was generally assumed that changes in high-level cognitive processes, such as decision-making, were responsible for improvements in perceptual performance with practice.

Various approaches, based largely on techniques in psychophysics and computational modeling, have been used in the study of perceptual learning. Psychophysics, which focuses on relationships between physical and sensory stimuli and mental processes, has provided especially useful insights into perceptual learning. Psychophysical techniques are designed to allow one to make inferences about the inner workings of a perceptual system by observing the responses that the system as a whole makes to carefully constructed stimuli. Psychophysical techniques have been used extensively to try to identify the kinds of cognitive processing changes that take place with practice in a wide variety of perceptual tasks.

Many psychophysical experiments have been applied to a wide array of tasks and stimuli involving other sensory modalities. Each of those applications is designed to uncover the underlying neural changes that take place with practice within a particular kind of perceptual processing. Examples of perceptual processes that have been investigated include visual motion detection, tactile spatial discrimination, and auditory frequency discrimination. Similar to Vernier acuity, for other sensory modalities there tends to be a high degree of specificity of learning with regard to task and stimulus, though there are important exceptions to that trend.

## 2.4 Infant-Directed Speech

A major component in infant language acquisition is interaction and communication. During infant interaction, a specific kind of language is used, often referred to as “baby talk”. Also known as Infant-Directed Speech, “baby talk” is the language spoken in an exaggerated rhythm or melody, therefore emphasizing word and phrase boundaries. Using IDS allows infants to more easily understand an adult’s stream of speech. This idea is demonstrated in Kuhl, Tsao and Liu's experiment in 2001. Nine-month-old American infants were exposed to a 5-hour play session where only Mandarin was spoken. These adults spoke in infant-directed speech, made eye contact with the child, and used the child’s name frequently. It was found that these children were as sensitive to Mandarin Chinese phonetic contrasts as infants whose native language was Mandarin. They were also significantly more sensitive to these contrasts than nine-month-old children who experienced the same five-hour play session but with an English speaking adult. This experiment is important because it was later compared to a similar experiment that used videotapes of Mandarin Chinese adults. At the end of the experiment the children who watched the Mandarin Chinese videos were tested for their sensitivity to contrasts in the language, and no significant difference was found between them and infants in the control group. Kuhl et al. used these two experiments to prove the importance of live interaction because of the social cues that help facilitate language acquisition.

# Chapter 3 Methodology

## 3.1 Research Questions

1. Which method is more effective for American Chinese learners to learn Chinese mandarin tones, the human-computer interaction based training software or the traditional way of learning through regular study materials?

2. How could the synthesized materials based on IDS help the learners to have a better categorical perception of Chinese mandarin tones?

## 3.2 Experiment Design

The experiment was designed and conducted to compare the capability of participants to identify and discriminate four Chinese Mandarin tones in two groups through an adaptive computer-based training program and self-designed ordinary training program. The language capability of participants was measured by pretest and post-test, where two sections, both an identification task and a discrimination task were presented to examine to what extents participants have improved their language ability to identify and discriminate Chinese Mandarin tones.

First, a preliminary training session was given, where participants were exposed with some basic knowledge about Chinese tones, pronunciations and tone diacritics. After giving the preliminary training session, all participants should be able to understand basic language rules about Chinese Mandarin tones and to have their own judgments about identification and discrimination of different Chinese Mandarin tones. Followed by the preliminary training session, a pre-test was presented to record participants' language capability of perceiving Chinese Mandarin tones before the training program, in which both an identification task section and a discrimination task were tested. After completing the pretest, 5 participants in experiment group 1 were exposed to an adaptive computer-based training program, where IDS (Infant-directed speech) was used as a key factor to give training to those mono-English speaking participants, while others in experiment group 2 were presented with a self-designed ordinary training program where participants received as same length of time as those in the experiment group 1.

In the experiment group 1, participants sit before the screen and receive Chinese Mandarin tone training with pairs in the following order: T1-T2, T1-T4, T2-T4, T1-T3, T3-T4, and T2-T3 (e.g. T2 represents tone 2, which is high rising tone). In the experiment group 2, participants sit in a quiet room and study by themselves with presented training materials. Both training programs lasted for about 2 hours, and all participants were presented with a post-test where the tasks remain similar with the pre-test, and scores were analyzed and evaluated to determine which training program is more effective and to what ways could the implications of experiment be utilized to improve productivity and effectiveness in second language learning. Furthermore, the outcome measures and data analysis section were presented to seek out the key features of Chinese Mandarin tones, and to identify the major difficulties for monolingual speakers of English. Also, feedback and reflections of participants regarding training program and the experiment were collected to help improve training program for a better experiment in the future. In the end, findings were demonstrated to point out the major results and key achievements to provide scholars with references and to illuminate further scientific exploration on the perceptual training of tones.

## 3.3 Participants

In this thesis, participants of the experiment were college students at University of Michigan Dearborn campus. Ten monolingual speakers of English were selected to participate the experiment. In order to make sure the accessibility of all participants could meet the candidates' qualification requirements of the experiment before the test, background check was implemented in several ways. First, all participants must be college students at the University of Michigan, where students demonstrated their language talents and learning capability at the same level. Second, all participants must be the students age from 18 to 26 in order to make sure all participants were in the same phase of life, where their neurotrophic factors, such as brain-derived neurotrophic factor (BDNF), insulin-like growth factor 1 (IGF-1) and vascular endothelial growth factor (VEGF) remain at the same level consistently (Comez-Pinilla, Hillman 2013). Third, all participants must never study Chinese Mandarin before to make sure the experiment were fair and effective, and participants' cognitive capability towards Chinese Mandarin tones were similar and perceptual training programs were able to function accurately and effectively. Only meet three criteria above, could participants allowed to access the experiment, and ten participants were selected strictly based on the three criteria to ensure the experiment could work functionally.

## 3.4 Preliminary Training Session

The preliminary training session is a simple training program before the pre-test, where all participants were provided with basic knowledge about Chinese Mandarin tones and tone diacritics. In this session, ten participants were presented with 5 examples for each tone with the same tone base ba, ma, shi, tian, yi. From Appendix A (Preliminary training material design), we could see the structure of preliminary training material. Each tone was presented with a tone name, tone diacritic, practice hint, and five examples with audios. For example, in second Mandarin tone- rising pitch, practice hint is "move your chin up when you practice the second tone", and examples are bá, má, shí, tián, yí. In third Mandarin tone- falling rising pitch, the practice hint is "move your chin down and then up again when you practice the third tone", and examples are bǎ, mǎ, shǐ, tiǎn, yǐ. Preliminary training session laid a solid foundation for the further pre-test and training program. Participants who received preliminary training should be able to have a clear understanding of Chinese Mandarin tone features and to make their own judgments in their identification and discrimination task.

## 3.5 Pre-test

A pre-test is a stage specially designed to test tone perceiving capability of participants prior to their training program. In this session, all ten participants in both experiment group 1 and experiment group 2 were instructed to take the same test in order to evaluate their cognition ability towards identification and discrimination of Chinese Mandarin tones.

The test was divided into two sections, and 30 questions in total were involved. In identification section, participants were provided with 20 identification questions which require the participants to identify and select the accurate tone of a pronounced given character from four possible tone choices. In discrimination section, participants were provided with 10 discrimination questions which presented in the form of tone combos where participants were required to discriminate the differentiation between the tone combos and select the correct one from all possible choices.

The test was presented and recorded by an online software (http://pinyinpractice.com/tones.htm) which was tested by the author to make sure it's reliable and effective to implement the pre-test. From Appendix B (Sample pages of software), we could discover the layout of the software and how it functions. The middle screen showed the character and its pinyin, participants could click on the character to listen to the pronunciation. Four tone choices were listed below the screen, and participants could select the correct one by clicking on the red dot. After selecting their answer, the system showed whether or not their answer is right, and record the answers on the right side of the column. Participants could proceed the test by clicking the next button on the lower right corner of the screen. The discrimination task was about the same, participants listened to the pronunciation of a tone combo, and make their choice by clicking on the white dots on the lower part of the screen, and the system automatically marked their answers and recorded on the right side of the column.

In order to examine the reliability of the online software (http://pinyinpractice.com/tones.htm), author tested 100 questions from the database of the software and record the occurrence rate of each tone. Through testing the accessibility and reliability of pre-test software, we calculated the occurrence rate of each tone, where tone 1 represents 27 percent of entire 100 questions while tone 2 represents 24 percent. Tone 3 makes up 23 percent, and tone 4 has 26 percent. The standard deviation (SD) = $\sqrt{\frac{1}{N}\sum_{i=1}^{N}(x_i-\mu)^2}$ = 1.58. This indicated that the variation of each number to the mean was small, namely, each tone has about the same probability to be involved into the pre-test, suggesting the software was reliable and effective to be utilized in the pre-test.

## 3.6 Training Program

The training program is a session where participants receive perceptual training towards Chinese Mandarin tones to enhance their performance to identify and discriminate four tones. The training program in the experiment group 1 is different from that in the experiment group 2. In the experiment group 1, an adaptive computer-based training software was presented to the participants and instructed them to train themselves in a certain method. However, in the experiment group 2, there's not a specific training session but a combination of various training methods and materials including videos, lecture, textbooks, and notes etc. based on a specially designed survey towards studying methods of second language learning. By comparing the differences of two training program, we could review the adaptive computer-based training program as an advanced, technology involved, double way transmission of knowledge based on human and computer interaction. In the meanwhile, the training program in experiment group 2 could be seen as an ordinary, traditional training method which is broadly ranged, one-way transmission of knowledge based on the experience and summary of instructors. Thus, the differences between two training programs are the key factors that make the outcomes varies. A deeper look at two training programs involved illustration of the design process, implementation process, features, and characteristic analysis.

### 3.6.1 Adaptive computer-based training software

The adaptive computer-based training software is a tone training software offered by author's instructor Professor Cheng. In this training session, participants received approximately 2 hours training through interaction with the software and were provided with 6 pair's comparative training modules in terms of T1-T2, T1-T4, T2-T4, T1-T3, T3-T4, and T2-T3. In each module, participants focused on specific training of two tones including identification of tones, comparison of two tones and discrimination of the tones. For example, module T2-T3 represents specific training of tone 2 and tone 3, participants were training targeting on the two tones, and only reached 90% accurate rate, could participants allowed to proceed to the next module.

### 3.6.2 Operation process

Each module of the software had a log in button, which could be clicked to get access into the training interface, and only IE browser could be used to enter the software, other browsers will be denied. Before logging into the software, participants' ID and names needed to be input to the software. By clicking the button Csexp, we could access to the back end of the software, where we could input participant's ID and names through double clicking on the participant button. The results of seven levels will also be recorded in this interface. After inputting participants' background information, we could click on the log in button to officially start the training session. In the trainee login interface, participants need to input their ID and name initials to enter the tone training interface. In the left upper corner of the interface, level number was indicated, while in the right lower corner, words listened were presented. There were 15 pronunciation of each tone, and when participant clicked on the tone icon, one character was pronounced in that tone. When completed the training, participants proceeded to take a test about the training they just received, and only reached 90 score, could participants go into next level. There were four speakers included in each module, and the number of speakers increased with enhancement of level of training. Practice hint followed with each tone set, and participants were encouraged to follow the hint to proceed their training session.

From Appendix C (The screen shots of training software), we could understand how the software operates. The entire training session consisted of 6 training module, and each module had 7 levels corresponding with 7 tests. Take module T1-T2 as an example. By clicking into the file T12, we could see it had 7 levels, in which each level followed with a test. There's were 7 levels and 7 tests in total. When participants received training of each level, they were instructed to proceed into the test with regard to the level they accomplished. After completing the test, if the accurate rate was more than 90 percent, then participants were allowed to enter to next level, otherwise, they had to repeat the level one more time. After repeating the same level, no matter what accurate rate was (reached 90 percent or not), participants were able to access to the next level. In other words, participants could repeat the same level twice maximum. From Fig. 3-1, Tone 1-2 module operation flow, we could have a precise look at the mechanism of operation of training software flow.

Fig. 3-1. Tone 1-2 module operation flow

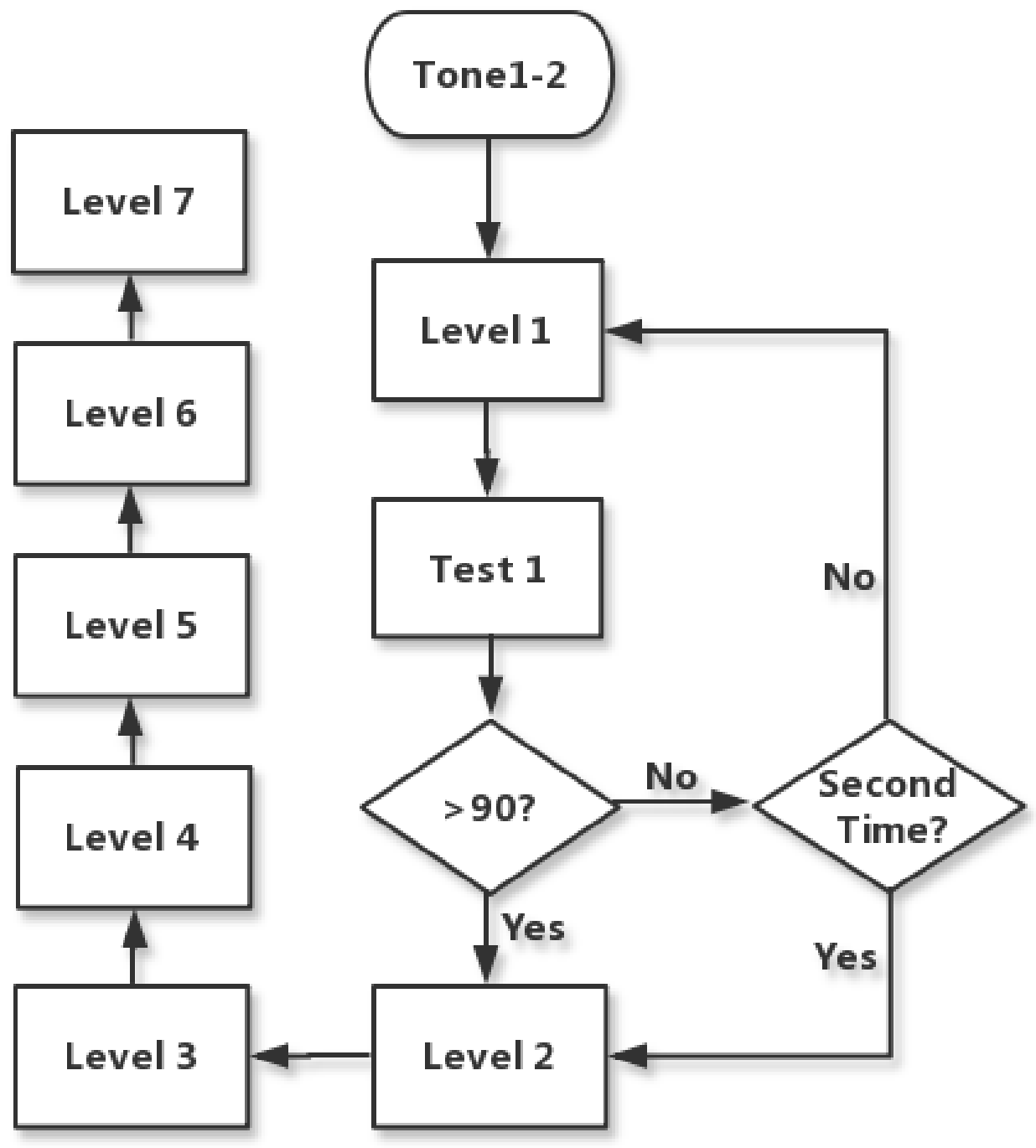

Tone 1-2 module operation flow represents how the training program functions. In module Tone 1-2, participants first reach to level 1. After completing the training in level 1, they were instructed to take test 1, if accuracy over 90%, they could get access to level 2, if below 90%, they have to repeat level 1 once again to move to the next level.

### 3.6.3 Self-designed training material

The training program in the experiment group 2 utilized self-designed training materials which consisted of the combination of various popular training materials including videos, lecture, textbooks, and notes based on the survey towards methods of second language learning. In this survey, as could be seen in Appendix D (A survey about traditional second language training material), three simple questions were asked to determine the most appropriate way to design the training material. The first question is "Have you ever learned a second language?", and interviewees could check either of two of the boxes indicating yes or no. Second questions is a multiple choice question which required the interviewees to select the most effective material they believe from the following four options: lectures, videos, textbooks, and notes. The third question asked the interviewees to list additional training materials which they believed are more effective than the given four options. From Fig. 3-2, the percentage of we could understand the proportion of "the most effective training materials", and determine how to distribute the various materials in the fixed 2-hour training session. Interviewees were encouraged to select multiple options if they feel it's necessary.

Fig. 3-2 Proportion of training material

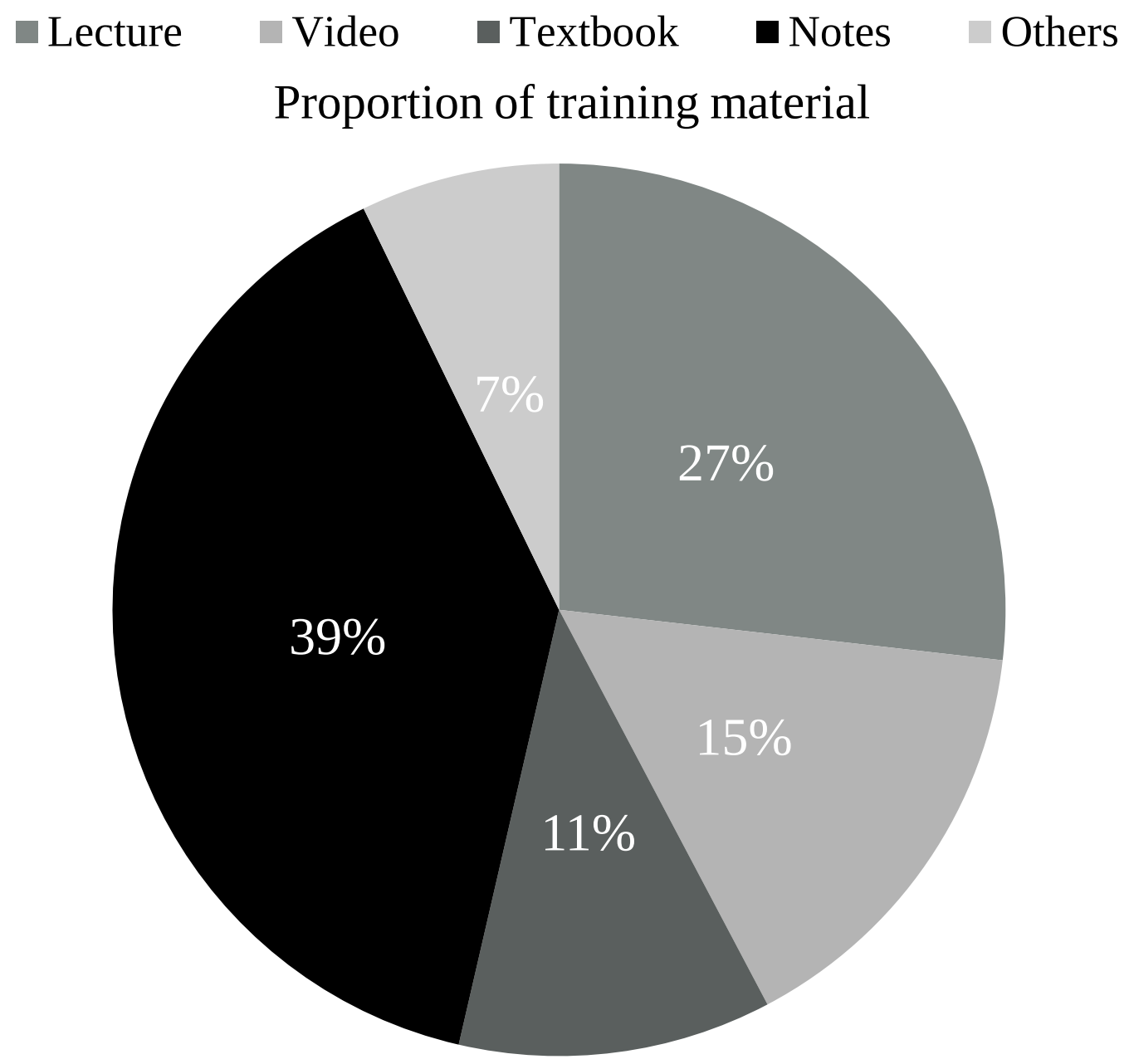

Proportion of training material represents the proportion of each training material in the training program in experiment group 2. For example, lecture makes up 27 percent of all material, then the length of time for lectures in the training program are 32 minutes. Notes make up 39 percent indicating the length of time for notes will be 46 minutes.

The textbooks are “Chinese in 10 minutes a day” (Second Revised edition) by Kristine K Kershul and “Beginners’ Chinese” with 2 audio CDs (Second Edition) by Yong Ho. In the meanwhile, two lectures and three videos were downloaded from YouTube channels where the addresses were showed as following.

1. https://www.youtube.com/watch?v=3wV8B4bx1lM

2. https://www.youtube.com/watch?v=5HjfI0n7JIM

3. https://www.youtube.com/watch?v=10p2AHD9hmE

4. +https://www.youtube.com/watch?v=1XQ7MwBmtKQ

5. https://www.youtube.com/watch?v=5-_P_H9gMmo

As for the notes, four notes were used in the training program for experiment group 2. One of them was an online note posted on the blog http://www.chinese-ilab.com/, and three of them were notes borrowed from students in Course ANTH 101 Introduction to Chinese LAN 103 Chinese oral communication, and ELP 201 Elementary Chinese 2.

Please note, it's not mandatory for participants in experiment group 2 to follow the proportion strictly. They were provided excess materials in every category, and were encouraged to select any material they feel comfortable. In this way, participants in experiment group 2 could make self-directed learning, and receive the most appropriate training by themselves.

## 3.7 Post test

The post-test is the session to evaluate the progress of training program. Participants were instructed to take an identical test with pretest including 20 identification task and 10 discrimination task. The contents of the test were not involved in previous training session and pre-test. Participants utilized same online software (http://pinyinpractice.com/tones.htm), and their scores were recorded and analyzed.

# Chapter 4 Results

## 4.1 Experiment Group 1

In the experiment group 1, a pre-test and a post-test were implemented to evaluate the effectiveness of perceptual training programs. In each test, 20 identification tasks and 10 discrimination tasks were presented. Thus, four sections in total were involved in the experiment group 1 in terms of Pre-stimuli in EXP-ID section (see Tab 4-1), Pre-stimuli in EXP-DS section (see Tab 4-2), Post-stimuli in EXP-ID section (see Tab 5-3), and Post-stimuli in EXP-DS section (see Tab 5-4). The thesis will present the results of each section, and discuss in detail about the meaning of the results.

### 4.1.1 Pre-stimuli in EXP1-ID

In this section, 5 participants were provided with 20 identification tasks, which require the participants to identify the correct one from four possible choices. From Tab 4-1, we could see the results of the section. The first column demonstrated the section name and 20 stimuli, and each stimulus contained with a character and its pinyin. In the second column, the standard tones of 20 stimuli were presented to conveniently compare with participants' answers. The third, fourth, fifth, sixth and seventh columns indicated the answers of five participants respectively. By reviewing their answers, we could notice that among 20 stimuli in the section, participants 1 chose eleven correct answers, and participants 2 selected nine right answers. While there are twelve accurate questions in participants 3's test, eleven correct answers were made by participant 4. Participant 5 had eight right answers. The accurate options were marked with a star to make reader easier to identify them and compare the wrong answers with standard tones in the second column.

Tab 4-1. Pre-Stimuli in EXP1-ID Group

| Pre- Stimuli in ***EXP1-ID*** Group | Standard Tone | Answer of Participant 1 | Answer of Participant 2 | Answer of Participant 3 | Answer of Participant 4 | Answer of Participant 5 |
|---|---|---|---|---|---|---|
| qiang<br>强 T 2 | T 2 | T 2* | T 3 | T 2* | T 2* | T 2* |
| zhang<br>张 T 1 | T 1 | T 1* | T 1* | T 4 | T 1* | T 4 |
| qi<br>气 T 4 | T 4 | T 4* | T 4* | T 1 | T 4* | T 1 |
| huang<br>黄 T 2 | T 2 | T 2* | T 1 | T 2* | T 1 | T 2* |
| jin<br>金 T 1 | T 1 | T 1* | T 1* | T 2 | T 3 | T 2 |
| qu<br>去 T 4 | T 4 | T 1 | T 4* | T 4* | T 4* | T 3 |
| lian<br>连 T 2 | T 2 | T 2* | T 2* | T 2* | T 2* | T 3 |
| wai<br>外 T 4 | T 4 | T 2 | T 1 | T 4* | T 2 | T 1 |
| ping<br>瓶 T 2 | T 2 | T 1 | T 2* | T 1 | T 3 | T 2* |
| si<br>死 T 3 | T 3 | T 3* | T 4 | T 3* | T 3* | T 1 |
| bei<br>北 T 3 | T 3 | T 3* | T 1 | T 3* | T 3* | T 3* |
| wu<br>五 T 3 | T 3 | T 2 | T 2 | T 3* | T 2 | T 3* |
| che<br>车 T 1 | T 1 | T 4 | T 4 | T 1* | T 4 | T 1 |
| cao<br>草 T 3 | T 3 | T 3* | T 3* | T 3* | T 4 | T 1 |
| huan<br>欢 T 1 | T 1 | T 2 | T 1* | T 2 | T 1* | T 3 |
| xie<br>谢 T 4 | T 4 | T 3 | T 1 | T 1 | T 4* | T 2 |
| xin<br>心 T 1 | T 1 | T 2 | T 4 | T 1* | T 1* | T 2 |
| lu<br>路 T 4 | T 4 | T 1 | T 2 | T 4* | T 3 | T 4* |
| bo<br>博 T 2 | T 2 | T 2* | T 4 | T 1 | T 2* | T 2* |
| dong<br>东 T 1 | T 1 | T 1* | T 1* | T 3 | T 4 | T 1* |
| Correct Number | 20 | 11 | 9 | 12 | 11 | 8 |

Pre-Stimuli in EXP1-ID Group represents the results of identification tasks in the experiment group 1in pretest. 20 identification questions were presented in the test, and participants have to identify the accurate one from the four options. The answer with an asterisk* on the upper right corner indicating this answer is correct. The answers without an asterisk * means it's wrong, and readers could figure out in which tone had participants misidentified.

### 4.1.2 Pre-Stimuli in EXP1-DS

In this section, 5 participants were provided with 10 discrimination tasks, which require the participants to discriminate the correct tone combo from all possible choices. From Tab 4-2, we could see the results of the section. The first column demonstrated the section name and 10 discrimination stimulus, and each stimulus contained with a character combo consisted of 2 characters and its pinyin. In the second column, the standard tones of 10 stimuli were presented to conveniently compare with participants' answers. The third, fourth, fifth, sixth and seventh columns indicated the answers of five participants respectively. By reviewing their answers, we could notice that among 10 tone combos in the section, participants 1and participant 2 selected four correct tone combos, and the rest of participants, namely, participant 2, 3, and 4 all scored three among 10 stimuli.

Tab 4-2. Pre-Stimuli in EXP1-DS Group

| Pre- Stimuli in ***EXP1-DS*** Group | Standard Tone | Answer of Participant 1 | Answer of Participant 2 | Answer of Participant 3 | Answer of Participant 4 | Answer of Participant 5 |
|---|---|---|---|---|---|---|
| shi cha 时差 | 2-1 | 4-2 | 2-1* | 2-1* | 2-4 | 2-1* |
| mei guo 美国 | 4-2 | 4-2* | 4-2* | 4-2 | 4-2* | 4-1 |
| bei jing 北京 | 4-1 | 1-1 | 1-4 | 4-1* | 4-2 | 4-1* |
| sui ran 虽然 | 1-2 | 1-2* | 4-2 | 1-3 | 1-2* | 4-2 |
| chu ban 出版 | 1-3 | 4-2 | 4-4 | 4-2 | 1-3* | 2-3 |
| kai hui 开会 | 1-4 | 4-4 | 4-1 | 1-2 | 4-3 | 4-4 |
| chu qu 出去 | 1-4 | 1-4* | 4-4 | 2-4 | 2-4 | 1-4* |
| kai shi 开始 | 1-3 | 4-2 | 4-1 | 2-4 | 4-4 | 1-3* |
| shang hai 上海 | 4-3 | 4-3* | 4-1 | 4-2 | 2-3 | 1-4 |
| zheng chang 正常 | 4-2 | 1-2 | 4-2* | 4-2* | 1-2 | 1-2 |
| Correct Number | 10 | 4 | 3 | 3 | 3 | 4 |

Pre-Stimuli in EXP1-DS Group represents the results of discrimination tasks in the experiment group 1 in pre-test. 10 discrimination questions were presented in the test, and participants have to identify the accurate tone combo from all possible options. The answer with a star mark* on the upper right corner indicating this answer is correct. The answers without a star mark* means it's wrong, and readers could figure out in which tone combo had participants misidentified.

### 4.1.3 Post-Stimuli in EXP1-ID

In this section, 5 participants in experiment group 1 were provided with a brand new test after accomplishing the training program, which neither of the stimuli was involved in the previous training or test session. Due to the training program, we expected 5 participants could enhance their accuracy to some extent, demonstrating the effectiveness of the training program. Through reviewing the data (as could be seen in Tab 4-3), all participants enhanced their accuracy to some points (except participant 3 remained same) just as we hoped. Participant 1 enhanced from eleven to fifteen, while participant 2 improved from nine to sixteen. Participant 3 remained the same to be eight, and participant 4 progressed from eight to fourteen. This phenomenon illustrated that the training program was indeed effective.

### 4.1.4 Post-Stimuli in EXP1-DS

The format of the post-test was identical to that in pre-test, however, the contents of the test changed dramatically different, and we expected participants were able to make progress in the discrimination tasks and have higher accuracy in comparison with the pre-test due to the training of the software. To make sure the test could work reliably and solidly, all repeated stimulus were deleted, and neither of the stimuli in this section was identical with that in others. By comparing Tab 4-2 with Tab 4-4 (showed on next page), we found out that Participant 1 had the accuracy of 8 out of 10, increasing about 4 points. Participant 2 increase his accuracy from 3 out of 10 to 5 out of 10. Participant 3 improved 4 points from 3 to 7, and participants 4 had improved most to be 6 points from 3 out of 10 to 9 out of 10. The participant 5 improved his accuracy to 6 out of 10. Through the simple comparison between the pretest and post-test in experiment group 1, the demonstration was clear that participants made progress in general in both identification task and discrimination task.

Tab 4-3. Post-Stimuli in EXP-ID Group

| Post- Stimuli in ***EXP1-ID*** Group | Standard Tone | Answer of Participant 1 | Answer of Participant 2 | Answer of Participant 3 | Answer of Participant 4 | Answer of Participant 5 |
|---|---|---|---|---|---|---|
| cheng 成 T 2 | T 2 | T 2* | T 2* | T 2* | T 1 | T 2* |
| xi 西 T 1 | T 1 | T 1* | T 1* | T 1* | T 3 | T 1* |
| lv 绿 T 4 | T 4 | T 4* | T 4* | T 2 | T 4* | T 4* |
| jiang 讲 T 3 | T 3 | T 3* | T 3* | T 3* | T 2 | T 3* |
| yi 已 T 3 | T 3 | T 3* | T 3* | T 3* | T 3* | T 3* |
| er 而 T 2 | T 2 | T 2* | T 2* | T 2* | T 2* | T 2* |
| mang 忙 T 1 | T 1 | T 1* | T 1* | T 3 | T 1* | T 4 |
| zhuo 桌 T 1 | T 1 | T 4 | T 1* | T 1* | T 1* | T 1* |
| zou 走 T 3 | T 3 | T 3* | T 3* | T 2 | T 4 | T 2 |
| zeng 增 T 1 | T 1 | T 1* | T 1* | T 4 | T 1* | T 3 |
| yin 印 T 4 | T 4 | T 4* | T 1 | T 2 | T 2 | T 4* |
| gui 鬼 T 3 | T 3 | T 1 | T 3* | T 2 | T 3* | T 3* |
| yue 月 T 4 | T 4 | T 4* | T 1 | T 4* | T 4* | T 1 |
| ci 次 T 4 | T 4 | T 4* | T 4* | T 4* | T 4* | T 2 |
| xiang 香 T 1 | T 1 | T 2 | T 1* | T 1* | T 1 | T 1* |
| qian 前 T 2 | T 2 | T 2* | T 2* | T 4 | T 2* | T 4 |
| ben 本 T 3 | T 3 | T 1 | T 2 | T 3* | T 3* | T 3* |
| long 龙 T 2 | T 2 | T 2* | T 3 | T 2* | T 2* | T 2* |
| shi 士 T 4 | T 4 | T 4* | T 4* | T 4* | T 4* | T 4* |
| shan 山 T 1 | T 1 | T 4 | T 1* | T 3 | T 2 | T 1* |
| Correct Number | 20 | 15 | 16 | 12 | 13 | 14 |

Post-Stimuli in EXP-ID Group represents the results of 20 identification tasks in the experiment group 1 in post-test. Through comparison between the pre-test and post-test, we could observe enhancement of accuracy for each participant, indicating cognitive ability of participants have improved to some extent after receiving the training of an adaptive computer-based training software

Tab 4-4. Post-Stimuli in EXP1-DS Group

| Post- Stimuli in ***EXP1-DS*** Group | Standard Tone | Answer of Participant 1 | Answer of Participant 2 | Answer of Participant 3 | Answer of Participant 4 | Answer of Participant 5 |
|---|---|---|---|---|---|---|
| yin yue 音乐 | 1-4 | 1-4* | 1-4* | 2-4 | 1-4* | 4-4 |
| lv ke 旅客 | 4-4 | 4-2 | 4-4* | 4-1 | 4-4* | 4-4* |
| kai hua 开花 | 1-1 | 1-1* | 1-4 | 2-1 | 1-1* | 1-2 |
| jing zhi 精致 | 1-4 | 1-4* | 1-4* | 1-4* | 1-4* | 1-4* |
| yan lei 眼泪 | 4-4 | 4-4* | 2-4 | 4-4* | 4-4* | 1-4 |
| jie shi 结识 | 2-2 | 2-2* | 2-1 | 2-2* | 2-2* | 1-2 |
| wang qiu 网球 | 4-2 | 4-2* | 4-1 | 4-2* | 4-2* | 4-2* |
| qi te 奇特 | 2-4 | 2-4* | 2-4* | 2-4* | 4-1 | 2-4* |
| chu fa 处罚 | 4-2 | 4-2 | 4-2* | 4-2* | 4-2* | 4-2* |
| xi qi 希奇 | 1-2 | 1-2* | 2-1 | 1-2* | 1-2* | 1-2* |
| Correct Number | 10 | 8 | 5 | 7 | 9 | 6 |

Post-Stimuli in EXP1-DS Group represents the results of 10 discrimination tasks in the experiment group 1 in post-test. The contents of the test are different from that in the pre -test, and participants also have to discriminate from given tone combos and select the correct answer from all possible options. Through comparison between Tab 4-2 and Tab 4-4, enhancement of accuracy in discrimination task was observed.

## 4.2 Experiment group 2

### 4.2.1 Pre-Stimuli in Exp2-ID

In this section, 5 participants in experiment group 2 were provided with the same test contents as that in experiment group 1. By reviewing the results (as could be seen in Tab 4-5), both participant 1 and 4 reached the accuracy of 14 out of 20, and participant 1 and 5 are about the same to be 7, and 8 respectively. Participant 3 had the accuracy of 10 out of 20. Because the experiment is randomized controlled, participants in both experiment group 1 and experiment group 2 ought to have similar cognitive ability before implementing the different training program. Thus, the results of the section should have little difference with that in EXP-ID section.

Tab 4-5. Pre-Stimuli in Exp2-ID Group

| Pre- Stimuli in ***Exp2-ID*** Group | Standard Tone | Answer of Participant 1 | Answer of Participant 2 | Answer of Participant 3 | Answer of Participant 4 | Answer of Participant 5 |
|---|---|---|---|---|---|---|
| qiang 强 T 2 | T 2 | T 1 | T 3 | T 2* | T 2* | T 2* |
| zhang 张 T 1 | T 1 | T 1* | T 2 | T 4 | T 1* | T 4 |
| qi 气 T 4 | T 4 | T 2 | T 2 | T 1 | T 4* | T 1 |
| huang 黄 T 2 | T 2 | T 2* | T 2* | T 2* | T 2* | T 2* |
| jin 金 T 1 | T 1 | T 1* | T 1* | T 2 | T 3 | T 2 |
| qu 去 T 4 | T 4 | T 1 | T 4* | T 4* | T 4* | T 3 |
| lian 连 T 2 | T 2 | T 2* | T 2* | T 2* | T 2* | T 3 |
| wai 外 T 4 | T 4 | T 4* | T 1 | T 4* | T 2 | T 1 |
| ping 瓶 T 2 | T 2 | T 2* | T 2* | T 1 | T 2* | T 1 |
| si 死 T 3 | T 3 | T 3* | T 4 | T 3* | T 3* | T 1 |
| bei 北 T 3 | T 3 | T 3* | T 1 | T 2 | T 3* | T 2 |
| wu 五 T 3 | T 3 | T 3* | T 3* | T 3* | T 2 | T 3* |
| che 车 T 1 | T 1 | T 4 | T 1* | T 1* | T 1* | T 1 |
| cao 草 T 3 | T 3 | T 3* | T 1 | T 2 | T 4 | T 3* |
| huan 欢 T 1 | T 1 | T 2 | T 1* | T 2 | T 2 | T 3 |
| xie 谢 T 4 | T 4 | T 4* | T 1 | T 1 | T 4* | T 4* |
| xin 心 T 1 | T 1 | T 1* | T 4 | T 1* | T 1* | T 2 |
| lu 路 T 4 | T 4 | T 1 | T 2 | T 4* | T 4* | T 1 |
| bo 博 T 2 | T 2 | T 2* | T 4 | T 1 | T 2* | T 2* |
| dong 东 T 1 | T 1 | T 1* | T 2 | T 3 | T 4 | T 1* |
| Correct Number | 20 | 14 | 8 | 10 | 14 | 7 |

Pre-Stimuli in Exp2-ID Group represents the results of 20 identification tasks in the experiment group 2 in pre-test. Because the experiment is randomized controlled, participants in both group ought to have similar cognitive ability before implementing the different training program. Thus, the results of the section should have little difference with that in EXP1-ID section.

### 4.2.2 Pre-Stimuli in Exp2-DS

In this section, 5 participants in experiment group 2 were provided with the same test contents as that in experiment group 1. By reviewing the results (showed in the Tab 4-6), both participant 1, 2 and 4 reached the accuracy of 3 out of 10, while the participant 3 made two correct answers and participant 5 selected five right choices.

Tab 4-6. Pre-Stimuli in Exp2-DS Group

| Pre- Stimuli in ***Exp2- DS*** Group | Standard Tone | Answer of Participant 1 | Answer of Participant 2 | Answer of Participant 3 | Answer of Participant 4 | Answer of Participant 5 |
|---|---|---|---|---|---|---|
| shi cha 时差 | 2-1 | 4-2 | 1-1 | 2-1* | 2-4 | 4-4 |
| mei guo 美国 | 4-2 | 1-2 | 4-2* | 4-2 | 2-2 | 4-2* |
| bei jing 北京 | 4-1 | 1-1 | 1-4 | 4-1* | 4-2 | 4-1* |
| sui ran 虽然 | 1-2 | 1-2* | 1-2* | 1-3 | 4-2 | 4-2 |
| chu ban 出版 | 1-3 | 1-3* | 4-4 | 4-2 | 4-2 | 1-3* |
| kai hui 开会 | 1-4 | 4-4 | 4-1 | 1-2 | 1-4* | 4-4 |
| chu qu 出去 | 1-4 | 1-4* | 1-4* | 2-4 | 1-4* | 2-4 |
| kai shi 开始 | 1-3 | 4-2 | 4-1 | 2-4 | 4-4 | 1-4 |
| shang hai 上海 | 4-3 | 1-2 | 4-1 | 4-2 | 4-3* | 4-3* |
| zheng chang 正常 | 4-2 | 1-2 | 1-2 | 4-2* | 1-2 | 4-2* |
| Correct Number | 10 | 3 | 3 | 2 | 3 | 5 |

Pre-Stimuli in Exp2-DS Group represents the results of 10 discrimination tasks in the experiment group 2 in pre-test. The contents of the test are identical to that in the experiment group 1, and participants have to discriminate between the tone combos from all possible options. Because the experiment is randomized controlled, participants in both group ought to have similar cognitive ability before implementing the different training program. Thus, the results of the section should have little difference with that in EXP-DS section.

### 4.2.3 Post-Stimuli in Exp2-ID Group

In this section, 5 participants in experiment group 2 were provided with the same test contents as that in EXP-ID group. Due to the training program, we expected 5 participants could enhance their accuracy to some extent. From Tab 4-7, we could see that participant 1 remained the same to be fourteen, while participant 2 enhanced from eight to fourteen. Participant 3 improved from ten to thirteen, and participant 4 progressed from seven to twelve. The results matched with our expectation.

Tab 4-7. Post-Stimuli in Exp2-ID Group

| Post- Stimuli in ***Exp2-ID*** Group | Standard Tone | Answer of Participant 1 | Answer of Participant 2 | Answer of Participant 3 | Answer of Participant 4 | Answer of Participant 5 |
|---|---|---|---|---|---|---|
| cheng 成 T 2 | T 2 | T 2* | T 2* | T 1 | T 2* | T 2* |
| xi 西 T 1 | T 1 | T 4 | T 1* | T 1* | T 3 | T 2 |
| lv 绿 T 4 | T 4 | T 3 | T 4* | T 4* | T 4* | T 4* |
| jiang 讲 T 3 | T 3 | T 3* | T 4 | T 3* | T 2 | T 3* |
| yi 已 T 3 | T 3 | T 3* | T 2 | T 3* | T 3* | T 1 |
| er 而 T 2 | T 2 | T 2* | T 2* | T 2* | T 1 | T 2* |
| mang 忙 T 2 | T 2 | T 2* | T 2* | T 3 | T 1 | T 2* |
| zhuo 桌 T 1 | T 1 | T 4 | T 1* | T 1* | T 1* | T 1* |
| zou 走 T 3 | T 3 | T 3* | T 3* | T 2 | T 4 | T 2 |
| zeng 增 T 1 | T 1 | T 4 | T 1* | T 1* | T 4 | T 3 |
| yin 印 T 4 | T 4 | T 4* | T 1 | T 2 | T 2 | T 4* |
| gui 鬼 T 3 | T 3 | T 1 | T 3* | T 3* | T 3* | T 3* |
| yue 月 T 4 | T 4 | T 4* | T 1 | T 3 | T 4* | T 4* |
| ci 次 T 4 | T 4 | T 4* | T 4* | T 4* | T 4* | T 2 |
| xiang 香 T 1 | T 1 | T 1* | T 2 | T 1* | T 1* | T 1* |
| qian 前 T 2 | T 2 | T 2* | T 1 | T 2* | T 2* | T 4 |
| ben 本 T 3 | T 3 | T 1 | T 3* | T 4 | T 3* | T 3* |
| long 龙 T 2 | T 2 | T 2* | T 2* | T 2* | T 2* | T 1 |
| shi 士 T 4 | T 4 | T 4* | T 4* | T 4* | T 2 | T 1 |
| shan 山 T 1 | T 1 | T 1* | T 1* | T 3 | T 2 | T 1* |
| Correct Number | 20 | 14 | 14 | 13 | 11 | 12 |

Post-Stimuli in Exp2-ID Group represents the results of 20 identification tasks in the experiment group 2 in the post-test. The results of the section showed an increase of accuracy in general.

### 4.2.4 Post-Stimuli in Exp2-DS

In this section, 5 participants in experiment group 2 were provided with the same test contents as that in EXP-DS after receiving traditional methods of training tones. We expected participants were able to make progress to some extent in the discrimination tasks due to the training session. However, the results ought to show variation with EXP-DS group since participants received totally different training modality. As indicated by Tab 4-8, participant 1 had the accuracy of 3 out of 10, remaining consistent with that in pre-test. Participant 2 increase his accuracy from 3 out of 10 to 5 out of 10. Participant 3 and 4 both increased 1 point from 2 to 3 and 3 to 4 respectively. The participant 5 remained his accuracy to be 5 out of 10. Through the simple comparison between the pretest and posttest in the experiment group 2, it could be seen that little progress had been made by participants in a discrimination task.

Tab 4-8. Post-Stimuli in Exp2-DS Group

| Post- Stimuli in ***Exp2-DS*** Group | Standard Tone | Answer of Participant 1 | Answer of Participant 2 | Answer of Participant 3 | Answer of Participant 4 | Answer of Participant 5 |
|---|---|---|---|---|---|---|
| yin yue 音乐 | 1-4 | 1-1 | 1-4* | 2-4 | 4-4 | 1-4* |
| lv ke 旅客 | 4-4 | 4-2 | 4-1 | 4-2 | 4-4* | 4-4* |
| kai hua 开花 | 1-1 | 1-1* | 2-2 | 1-4 | 1-1* | 1-4 |
| jing zhi 精致 | 1-4 | 1-2 | 1-1 | 1-4* | 4-4 | 2-4 |
| yan lei 眼泪 | 4-4 | 4-4 | 4-4* | 4-4 | 4-3 | 4-1 |
| jie shi 结识 | 2-2 | 4-2 | 2-1 | 2-2* | 2-3 | 4-2 |
| wang qiu 网球 | 4-2 | 4-2* | 2-3 | 4-2 | 4-2* | 4-4 |
| qi te 奇特 | 2-4 | 2-4* | 2-4* | 1-4 | 2-3 | 2-4* |
| chu fa 处罚 | 4-2 | 4-2 | 4-2* | 4-2* | 4-3 | 4-2* |
| xi qi 希奇 | 1-2 | 1-4 | 1-2* | 1-3 | 1-2* | 1-2* |
| Correct Number | 10 | 3 | 5 | 3 | 4 | 5 |

Post-Stimuli in Exp2-DS Group represents the results of 10 discrimination tasks in the experiment group 2 in the post-test. The contents of the test are identical to that in the experiment group 1, and participants have to discriminate the tone combos from all possible choices. The results demonstrated that little progress could be observed in discrimination task in experiment group 2.\

# Chapter 5 Findings

## 5.1 Outcome measures

### 5.1.1 Dispersion measures

In order to resolve first research question "Which training program is better and more effective and why?” dispersion measures were crucial to address the problem. In the experiment group 1, through calculation of increase accuracy (Tab 5-1), we found out that participant 2 had the highest increase rate, reaching to 30 percent, while participant 3 had least increase rate to be 13.33%. Participant 1, 4 and 5 had extremely close increase rate to be 26.67%, 26.66%, and 26.67% respectively. The average of accuracy increase rate showed to be 24.66%.

As for experiment group 2, participant 3 had the highest increase rate to be 26.67% while participant 3 and 5 had a close rate to be 13.33% and 16.67% in specific (Tab 5-2). What needs to be mentioned was participant 4 showed a slight recession with -6.67% increase rate, decreasing from 56.67% to 50%. Participant 1 remained the same rate of 56.67% before and after the training session. In additional, the mean increase rate was 10%.

Apparently, through comparison between Tab 5-1 and Tab 5-2, we could easily resolve the question and generate the conclusion that training program in experiment group 1 was better than that in the experiment group 2 for two following reason. First, the accuracy increase rate was pervasively higher in experiment group 1 with the mean of 24.66% than experiment group 2 with only 10%. Second, the accuracy increase rate for each individual in experiment group 1 was much more consistent than that in experiment group 2. When we look at the experiment group 1, each input was positive and standard deviation (SD) =6.5, and standard error of the mean (SE) =2.91. However in the experiment group 2, SD was 13.3 > 6.5, and SE was 5.96 > 2.91. Thus, measures of dispersion were much bigger in experiment group 2 than that in experiment group 1, suggesting the increase of accuracy in experiment group 1 was higher and more consistent. In this way, an adaptive computer-based training software was better and more effective due to higher accuracy increase, more consistent and lower measures of diversion.

Tab 5-1. Accuracy increase rate in experiment group 1

| Pre-Post | Test 1% | Test 2% | Test 3% | Test 4% | Test 5% | Mean |
|---|---|---|---|---|---|---|
| Pre- Exp1 | 50.00 | 40.00 | 50.00 | 46.67 | 40.00 | 45.33 |
| Post- Exp1 | 76.67 | 70.00 | 63.33 | 73.33 | 66.67 | 70.00 |
| Increase % | 26.67 | 30.00 | 13.33 | 26.66 | 26.67 | 24.66 |

Accuracy increase rate in experiment group 1, represents the increase rate of each individual, with the highest number 30%, lowest 13.33%, and mean 24.66%

Tab 5-2. Accuracy increase rate in experiment group 2

| Pre-Post | Test 1% | Test 2% | Test 3% | Test 4% | Test 5% | Mean |
|---|---|---|---|---|---|---|
| Pre- Exp2 | 56.67 | 36.67 | 40.00 | 56.67 | 40.00 | 46.00 |
| Post- Exp2 | 56.67 | 63.33 | 53.33 | 50.00 | 0.57 | 56.00 |
| Increase % | 0.00 | 26.67 | 13.33 | -6.67 | 16.67 | 10.00 |

Accuracy increase rate in experiment group 2, represents the increase rate of each individual, with the highest number 26.67%, lowest -6.67%, and mean 10%

### 5.1.2 ID-DS measures

In order to resolve the question 2 and 3, ID-DS measures needed to be conducted to figure out how the training programs functioned on identification task and discrimination task. As a matter of fact, the problem could easily be addressed through comparison between identification section and discrimination section (Tab 5-3). In Tab 5-3, four categories in terms of Exp1-ID, Exp1-DS, Exp2-ID, and Exp2-DS were listed in which the increase rates of pretest and post-test were calculated and presented. The increase rate of Exp1-ID was almost identical to that of Exp2-DS to be 19% and 18%. The accuracy increase in Exp2-ID was relatively lower to be only 11%. The focus point was Exp1-DS section, where the accuracy increase went dramatically high, reaching to 36%.

Tab. 5-3 ID-DS Comparison

| ID-DS (%) | Exp1-ID% | Exp1-DS% | Exp2-ID% | Exp2- DS% |
|---|---|---|---|---|
| Pre- test | 51.00 | 34.00 | 53.00 | 32.00 |
| Post- test | 70.00 | 70.00 | 64.00 | 40.00 |
| Increase % | 19.00 | 36.00 | 11.00 | 18.00 |

ID-DS Comparison represents the increase in four categories from pre-test to post-test. Exp means experiment group 1, ID means identification task, Exp2 means experiment group 2, and DS means discrimination task.

To further analyze the statistics and to have a straightforward perspective towards accuracy increase of four categories, a bar chart (Fig 5-1) and a line chart (Fig 5-2) was presented for readers in order to have a better understanding. In Fig 5-1, we noticed that the heights of Exp2-ID and Exp1-ID were about the same while Exp2-DS and Exp1-DS differed dramatically, indicating, the adaptive computer-based software and traditional training materials have about the same effects on identification task. However, in discrimination task, computer training software outweighed traditional materials tremendously.

In Fig 5-2, the trend of line was even more obvious for readers to comprehend. While Exp2-ID, Exp1-ID, and Exp2-DS struggled to vibrate in the spectrum ranging from 10 to 20, Exp1-DS increased quickly across two spectra from 20 to 40. Thus, thesis concluded that although experiment group 1 and experiment group 2 functioned in similar in identification task, the training program in experiment group 1 performed way better than that in experiment group 2.

Fig 5-1, Bar chart of ID-DS

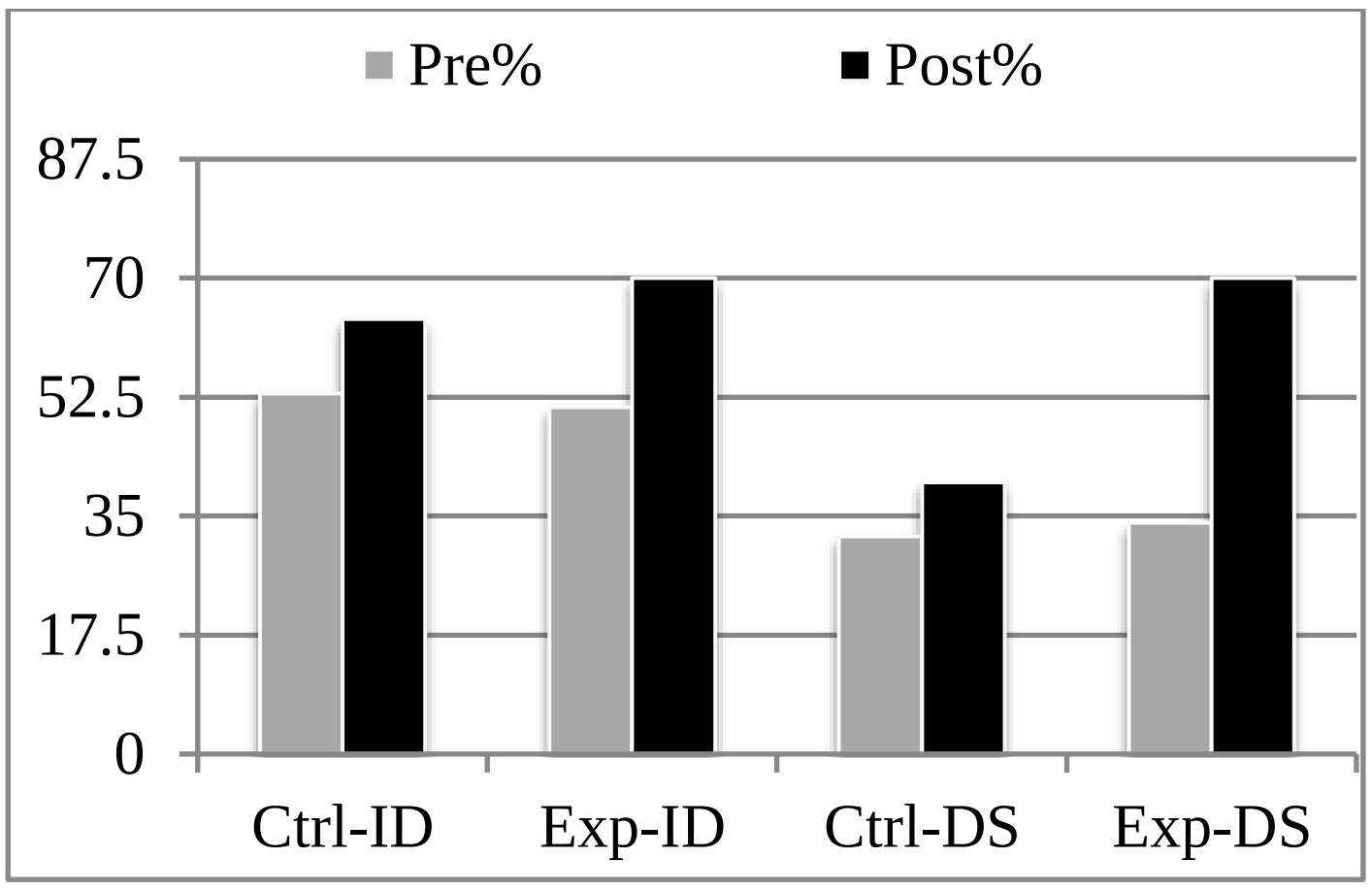

Bar chart of ID-DS, represents the increase rate from pre-test to post-test in four different categories in a straightforward way. Adaptive computer based training program functioned more effective in discrimination task

Fig 5-2, line chart of ID-DS

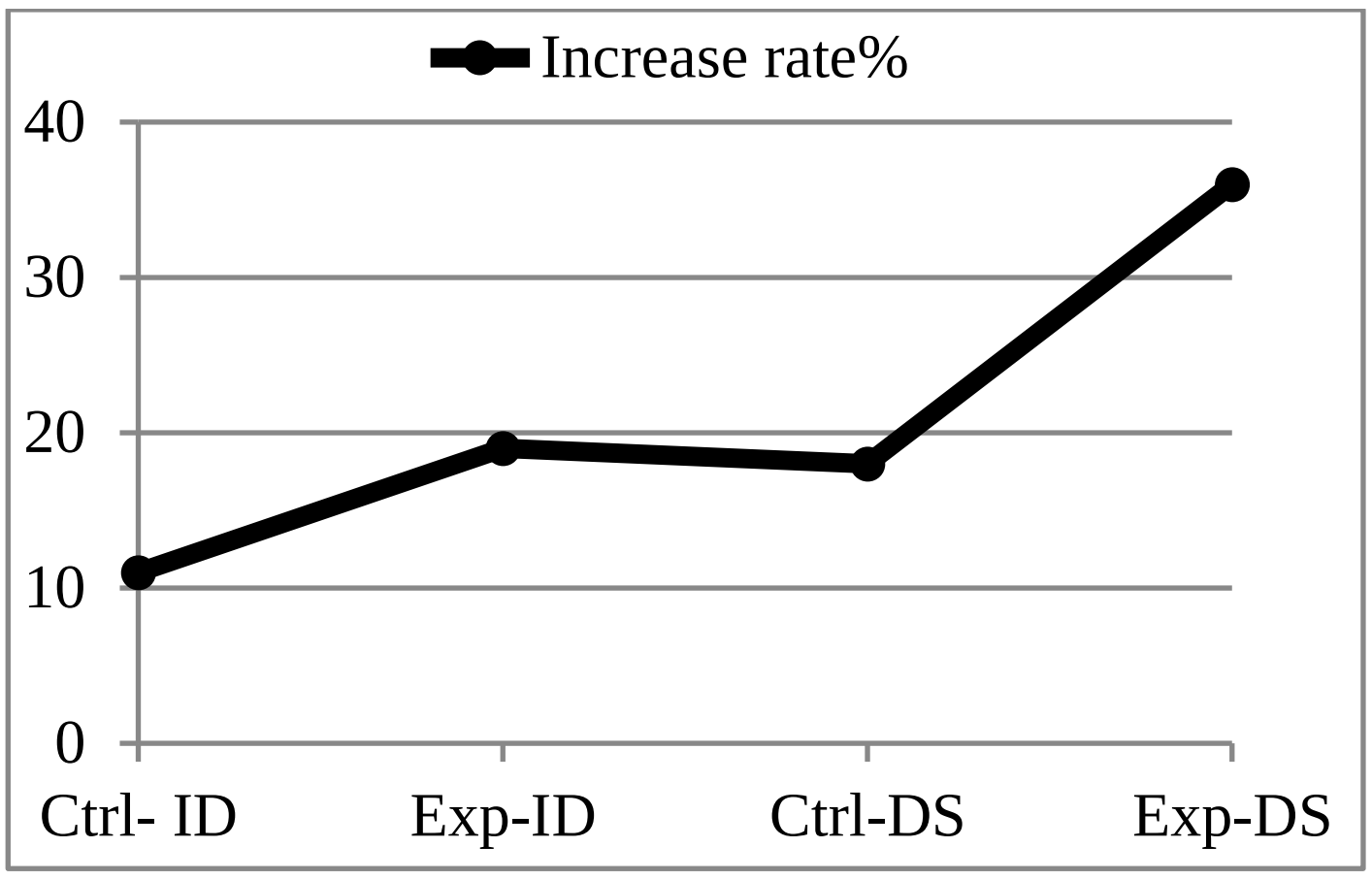

Line chart of ID-DS, represents the trend of increase rate among four categories. In identification section, experiment group 1 performed better than experiment group 2 ranging from 10 to 20, while in discrimination section, computer software outweighed traditional materials in a dramatic way.

### 5.1.3 Tone performance measures

In this section, thesis analyzed the results for each tone to measure the performance of training gains. In order to figure out the question 4 "How the training programs functioned on each tone?", comparisons of training gains in each tone and contrasts between identification task and discrimination task in both experiment group 1 and experiment group 2 were implemented to detect the findings behind the numbers. In experiment group 1 (Tab 5-4), tone 3 had the least training gains to be 8% and tone 4 had the most gains 32% in identification task while tone 2 had the least 23.3% and tone 4 also had the most gains 47%. In general, tone 2 improved the least to be 19.78% and tone 4 gained most being 39.5%. From tab 5-4, we can see tone performance of each tone in both tasks including percent of the correctness of pre-test and post-test. In the first column, tone names were listed and column 2, 3 and 4 indicating the outcomes in identification task in terms of percent correct pre, percent correct post and training gains. Column 5, 6, 7 represents the outcomes in discrimination task while column 8, 9, 10 summarized outcomes in total experiment group 1. By comparing the discrimination task and identification task, we observed that discrimination task had more training gains than that in experiment group1 with gains of each tone outweighed. For example, Tone 1 in DS task had 35.43% gains > 20% gains in ID task. This indicated that in experiment group1, the training program was more effective to discrimination task than identification task.

Tab. 5-4. Tone performance measure in Experiment group1

| EXP Group1 | Identification Task | | | Discrimination Task | | | Total Experiment Group1 | | |
|---|---|---|---|---|---|---|---|---|---|
| | Percent Correct Pre | Percent Correct Post | Training gains | Percent Correct Pre | Percent Correct Post | Training gains | Percent Correct Pre | Percent Correct Post | Training gains |
| Tone 1 | 43.33 | 63.33 | 20.00 | 28.57 | 64.00 | 35.43 | 35.95 | 63.67 | 27.72 |
| Tone 2 | 64.00 | 80.00 | 16.00 | 50.00 | 73.33 | 23.33 | 57.00 | 76.67 | 19.67 |
| Tone 3 | 60.00 | 68.00 | 8.00 | 32.00 | 70.00 | 38.00 | 46.00 | 69.00 | 23.00 |
| Tone 4 | 40.00 | 72.00 | 32.00 | 25.00 | 72.00 | 47.00 | 32.50 | 72.00 | 39.50 |

Tone performance measure in Experiment group1 represents the outcomes of training gains of each tone in both identification task and discrimination task. Three sections were included in the tab in terms of ID task, DS task and Total Experiment Group. By evaluating the data, we found that tone 4 improved most in all sections while tone 3 improved least 8% in ID task and tone 2 had least gains in both DS section(23.33%) and total group(19.67%).

In experiment group 2, (Tab 5-5), tone 2 and 3 had the least training gains to be 0% and tone 4 had the most gains 20% in identification task while tone 1, 2, and 3 had similar training gains around 8%, tone 4 also had -4% training gains. In general, tone 1 improved the most to be 12.62%, tone 2, 3 improved around 4% and tone 4 gained 8%. From Tab 5-5, we observed that there wasn't a strong correlation between ID and DS task.

Tab 5-5. Tone performance measures in Exp2 group

| Exp2 Group | Identification Task | | | Discrimination Task | | | Total experiment group 2 | | |
|---|---|---|---|---|---|---|---|---|---|
| | Percent Correct Pre | Percent Correct Post | Training gains | Percent Correct Pre | Percent Correct Post | Training gains | Percent Correct Pre | Percent Correct Post | Training gains |
| Tone 1 | 43.33 | 60.00 | 16.67 | 31.43 | 40.00 | 8.57 | 37.38 | 50.00 | 12.62 |
| Tone 2 | 72.00 | 72.00 | 0.00 | 35.00 | 43.33 | 8.33 | 53.50 | 57.67 | 4.17 |
| Tone 3 | 60.00 | 60.00 | 0.00 | 32.00 | 40.00 | 8.00 | 46.00 | 50.00 | 4.00 |
| Tone 4 | 44.00 | 64.00 | 20.00 | 40.00 | 36.00 | -4.00 | 42.00 | 50.00 | 8.00 |

Tone performance measures in Exp2 group represented the tone performance in experiment group 2. Through analysis of data in two tasks, we observed neither a correlation between gains in two tasks nor a trend among four tones. The numbers were unpredictable and uncontrollable.

1. For each tone, participants in experiment group1 had better training gains in general than those in experiment group 2, suggesting computer-based training program is more effective towards the perceptual training of Chinese tones than tradition training material.

2. By comparing the training gains of each tone in two tasks in both groups, experiment group1 had more consistent and positive effects on all tones while performances of tone training were more disperse in experiment group 2.

3. In experiment group1, Tone 4 had most training gains in both identification and discrimination task. While Tone 3 struggled to improve in identification task, Tone 2 had less impressive performance in discrimination task and the total group in general.

4. In experiment group1, participants had better performance on every tone in discrimination task than identification task, indicating computer based software was more effective in discrimination task training.

5. In experiment group 2, there was neither a significant correlation between two tasks nor a predictable trend among four tones.

## 5.2 Feature analysis

Why is the computer training software so special in its training procedure design? In what ways does the software differ from the traditional training materials? Why could the software function effectively towards the perceptual training of tones which were though only could be taught by teachers? In order to answer these questions, feature analysis of the software is crucial to understanding the key features of the software and the mechanism of adaptive computer-based training program.

1. Infant-directed speech (IDS) training modality: Zhang et al. (2009) suggested that the combination of characteristics of IDS and an adaptive computer-based training software could improve the effects of perceptual training of tones. According to Zhang (2011), IDS could exaggerate spectral and temporal cues for pitch and phonetic categories, and its adaptive modification based on learning.
2. Visual articulation cues: An icon indicating the status of levels sits in the upper left corner of the screen, while "words listened" is presented in the lower right corner of the interface. A static photograph of a speaker situated in the bottom left of the screen.

3. Self-paced training model: The participants selected the stimuli by clicking on the icon button of corresponding tone diacritic. For example, if participants click on the icon of rising tone (T 2), then the speaker pronounced the character with the second Mandarin tone. In this way, participants controlled their own behavior of making choice between the given tone pairs. Also, if the participants select the same tone for 30 times, the software will instruct the participants to select a different tone.
4. Variation of difficulties: The training for each module progressed through 7 levels of difficulties. The first level contained the most exaggerated pronunciation spoken by only one speaker. The second, third and fourth level remained the same exaggeration but spoken by one additional speaker with the progress of each level. The fifth and sixth remained all four speakers while decreased the exaggeration of pronunciation. In level seven, natural speeches were used in the training. Participants had to take a short quiz consisted of 5 identification questions of each tone randomly selected from a database of exemplars. Only reached 90 percent accuracy, can participants move on to the next level.
5. A wide range of exemplar with high variability: According to Heinzen (2014), there are 30 examples in each tone at each of the four stages of exaggeration. Thus, a total of 480 speeches were utilized in the training program, and they were spoken by four speakers including 2 males and 2 females. Participants could receive a comprehensive training towards Chinese Mandarin tones, in which they were exposed with four stages of exaggeration, four different pitch range and phonetic categories, and a wide range of exemplars.
6. Pair-focusing comparison: Y. Wang et al. (1999) suggested that adaptive scaffolding is a key feature of the training software, in which 6 modules trained from the least difficult contrasts to the most difficult contrasts in the following order: T1-T2, T1-T4, T2-T4, T1-T3, T3-T4, and T2-T3.
7. Human-computer interaction: The traditional tone training software utilized the teaching method of one-way transmission of knowledge, which instructors shared their knowledge and experience with trainees who was not able to interact with instructors often. However, adaptive computer-based training program utilized a more advanced method of teaching based on a cross-way transmission and human-computer interaction. In this way, trainees were self-directed and were able to interact with training materials much more often.

# Chapter 6 Conclusion

The thesis designed a randomized controlled experiment, selected 10 participants with similar accessibility, and distributed them into two groups. After offering the same preliminary training session for all participants, they were instructed to complete a pre-test, which included 20 identification questions and 10 discrimination questions based on an online software, evaluating their capability of identifying and discriminating four Chinese Mandarin tones. Participants in two groups were provided with different training modality. In experiment group1, an adaptive computer-based training software was utilized and trained participants in the order from least difficult contrasts to most difficult contrasts. In experiment group 2, traditional training program was designed based on the reflection of attitudes of surveys including lectures, videos, notes, textbooks and others. The length of the training programs were controlled within two to two and a quarter hours, where participants in two groups were exposed with different training modalities. After the training programs, participants in two groups took a post-test with was similar in the format with the pre-test. The results of pre-test and post-test were recorded, audited, analyzed and evaluated, and five research questions were proposed. In order to address the five research questions, outcome measures in terms of dispersion measures, ID-DS measures, tone performance measures, and feature analysis were implemented, and crucial findings and implications were summarized.

## 6.1 Major Findings

In dispersion measures, thesis analyzed accuracy increase in two groups, and discovered that participants in experiment group1 had better accuracy in the post-test than those in experiment group 2, and SD was smaller in Experiment group1 as well, indicating computer based training program was generally more effective, and had positive effects towards every participants. In experiment group 2, some participants showed accuracy increase in the post-test, however, some showed little improvement and even decrease in the post-test.

In ID-DS measures, thesis evaluated the performance targeting on ID and DS task, where four modules Exp1-ID, Exp1-DS, Exp2-ID and Exp2-DS, were compared. We observed that although there was a little increase between Exp2-ID and Exp1-ID, Exp1-DS was doubled than Exp2-DS, indicating computer based training software had significant effects on training discrimination between Chinese tones.

In tone performance measure, thesis analyzed and evaluated the performance of four tones in each task in two groups. Generally speaking, participants in Exp group1 had more accuracy for each tone than those in Exp group 2. Also, the accuracy increases were more consistent and sustainable for each tone in Exp group 1 which means every tone improved its accuracy in the post-test while Tone 2 and 3 showed no improve, and Tone 4 declined 4% in the Exp2 group. Specifically speaking, Tone 4 had the most training gains in both tasks, tone 3 had least increase in ID task and tone 2 had least in DS task. Furthermore, participants in Exp group1 performed better in DS task than ID task.

In feature analysis section, seven features were demonstrated to explain why an adaptive computer-based training software performed better than traditional materials. First, IDS and four speakers were contained in the software to exaggerate pitch range and phonetic categories. Second, visual articulation cues kept trainee to stay focused. Third, self-paced training model enabled trainee to adjust their training based on their own conditions. Fourth, 7 levels of difficulties were included in the software, which participant had to reach 90% accuracy to move to the next level. Fifth, a wide range of exemplars of 480 in total were contained in the software, which gave trainee detailed examples of different pitch range and phonetic categories. Sixth, pair-focused comparison was utilized to give training from the least difficult contrasts to the most difficult contrast. Seventh, participants were able to interact with software in various ways which enabled the trainee the absorb knowledge in the double way of transmission.

All in all, the adaptive computer-based training software outweighed traditional tone training materials in various performance, and it had a significant impact on perpetual training of discrimination task. Among four Chinese Mandarin tones, tone 4 improved the most, tone 3 increase the least in identification task and tone 2 had lowest accuracy in discrimination task. Seven features of software were analyzed, demonstrating the key factors that make computer training software more effective. The success of the experiment served as a testament to prove that AI and human-computer interaction could produce better performance on perceptual training of Chinese Mandarin tones, which not only discovered an innovative tone training modality, but also laid a solid foundation for further research about perceptual training of tones and human-computer interaction.

## 6.2 Implications

The study explored a new technique of phonetic tone training, which may have positive impact on second language learning and tone training. The experiment examined the feasibility and effectiveness of an adaptive human-computer interaction based tone training model employing infant-directed speech, visual cues and listening materials with variations of difficulties through the techniques of outcome measures and feature analysis. The major findings testified that human-computer interaction based training program has great potential in tone training and could be utilized in language education and phonetic tone training for second language learners in adulthood.

Furthermore, tone 2 and tone 3 have been examined to be the most difficult tone to differentiate with other tones and the hardest tone to identify respectively. This finding could illuminate Chinese Mandarin learners to give specifically emphasized training towards the two tones, and hence improve their Mandarin proficiency dramatically.

## 6.3 Limitations

Two limitations still existed in this thesis. First, the study lack of lab and sufficient funding to support the conduction of experiment. Since the study require a relatively acute observation and precise calculation of tone performance of each participant, the experiment place and instruments should be unified to eliminate confounding factors.

Second, the study lack of sufficient manpower to assist in the experiment. Since each participant lasted for 3 hours in the experiment, it’s difficult to handle all the operation by only one experimenter.

# Appendix A: Preliminary Training Materials

Look at the graphic below. It represents tones within one's relative vocal range.

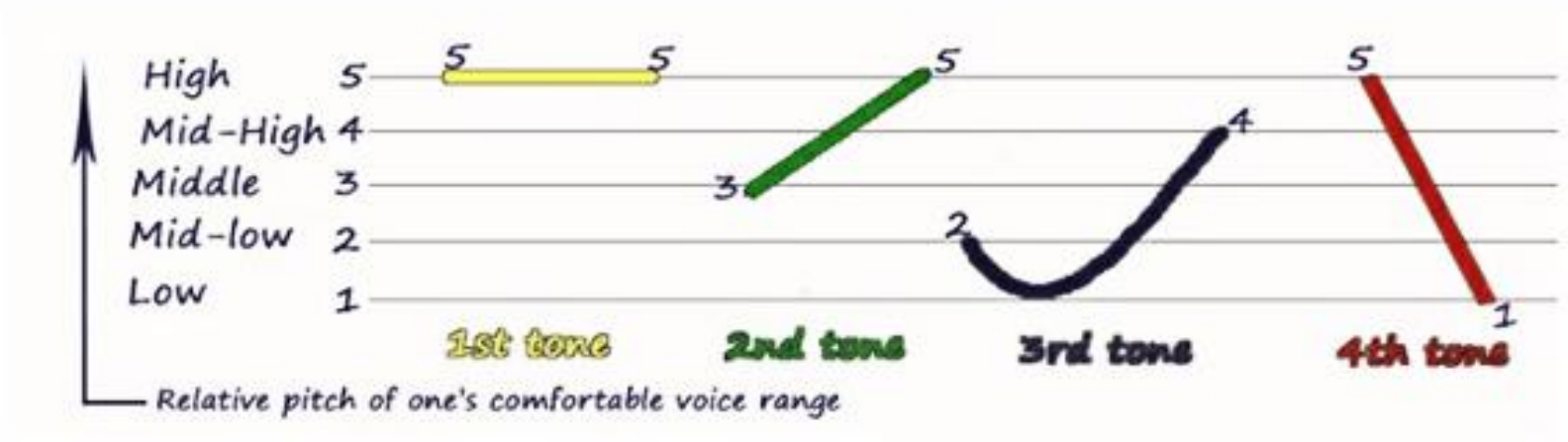

**Practice hint:** Move your head from left to right when you pronounce syllables with the first tone.

Now, listen to the examples below and try to repeat them at your upper limit

| Pinyin | bā | mā | shī | tiān | yī |
|---|---|---|---|---|---|
| Audio | ▶ | ▶ | ▶ | ▶ | ▶ |

**Practice Hint:** Move your chin up when you practice the second tone.

| Pinyin | bá | má | shí | tián | yí |
|---|---|---|---|---|---|
| Audio | ▶ | ▶ | ▶ | ▶ | ▶ |

**Practice hint:** Move your chin down and then up again when you practice the third tone.

Listen to the examples and test your voice at the lower vocal range to get used to the sound.

| Pinyin | bǎ | mǎ | shǐ | tiǎn | yǐ |
|---|---|---|---|---|---|
| Audio | ▶ | ▶ | ▶ | ▶ | ▶ |

**Practice hint:** Move your head down as if you were nodding when you practice the fourth tone.

| Pinyin | bà | mà | shì | tiàn | yì |
|---|---|---|---|---|---|
| Audio | ▶ | ▶ | ▶ | ▶ | ▶ |

# Appendix B: Sample pages of Online Testing Software

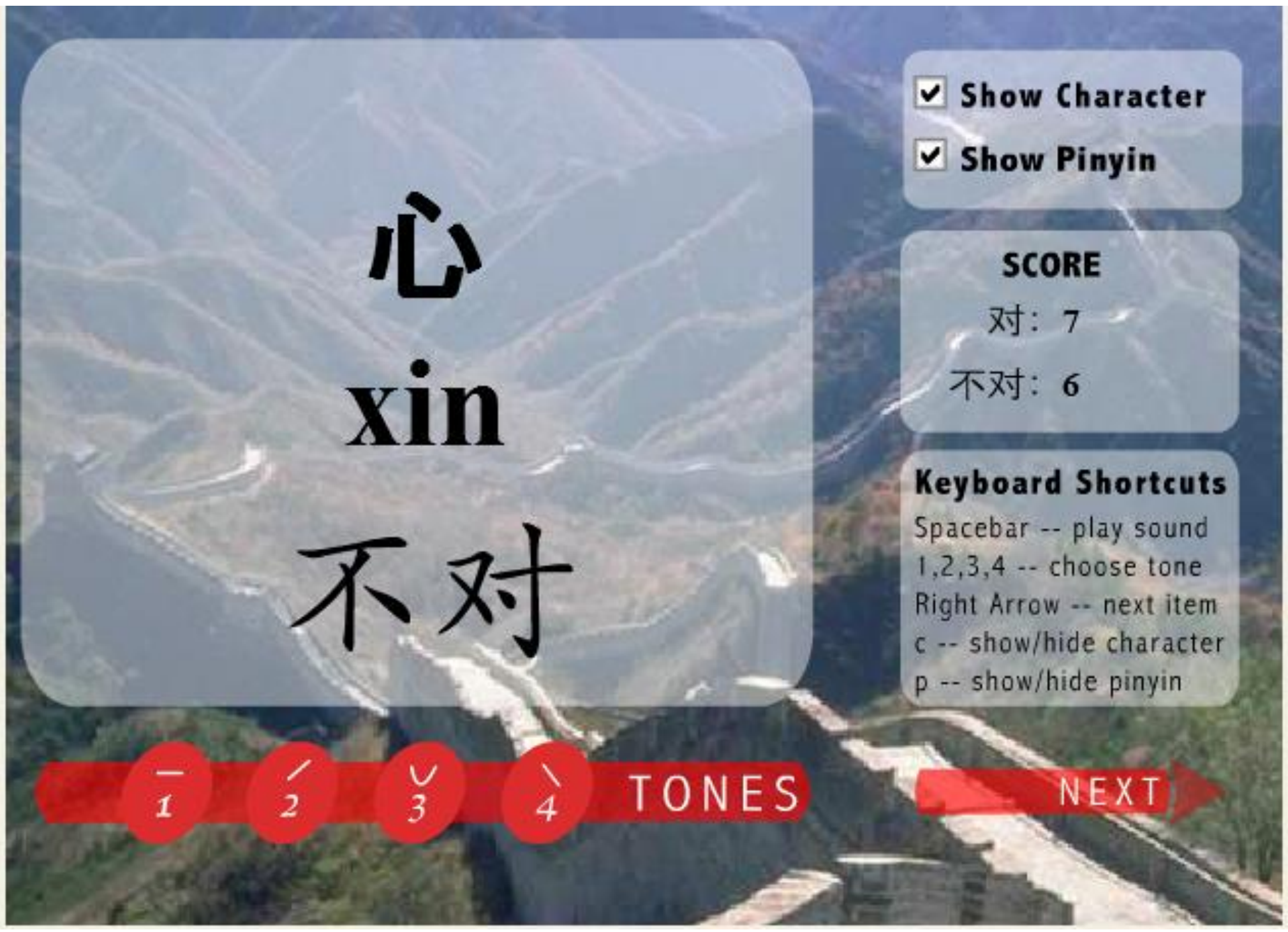

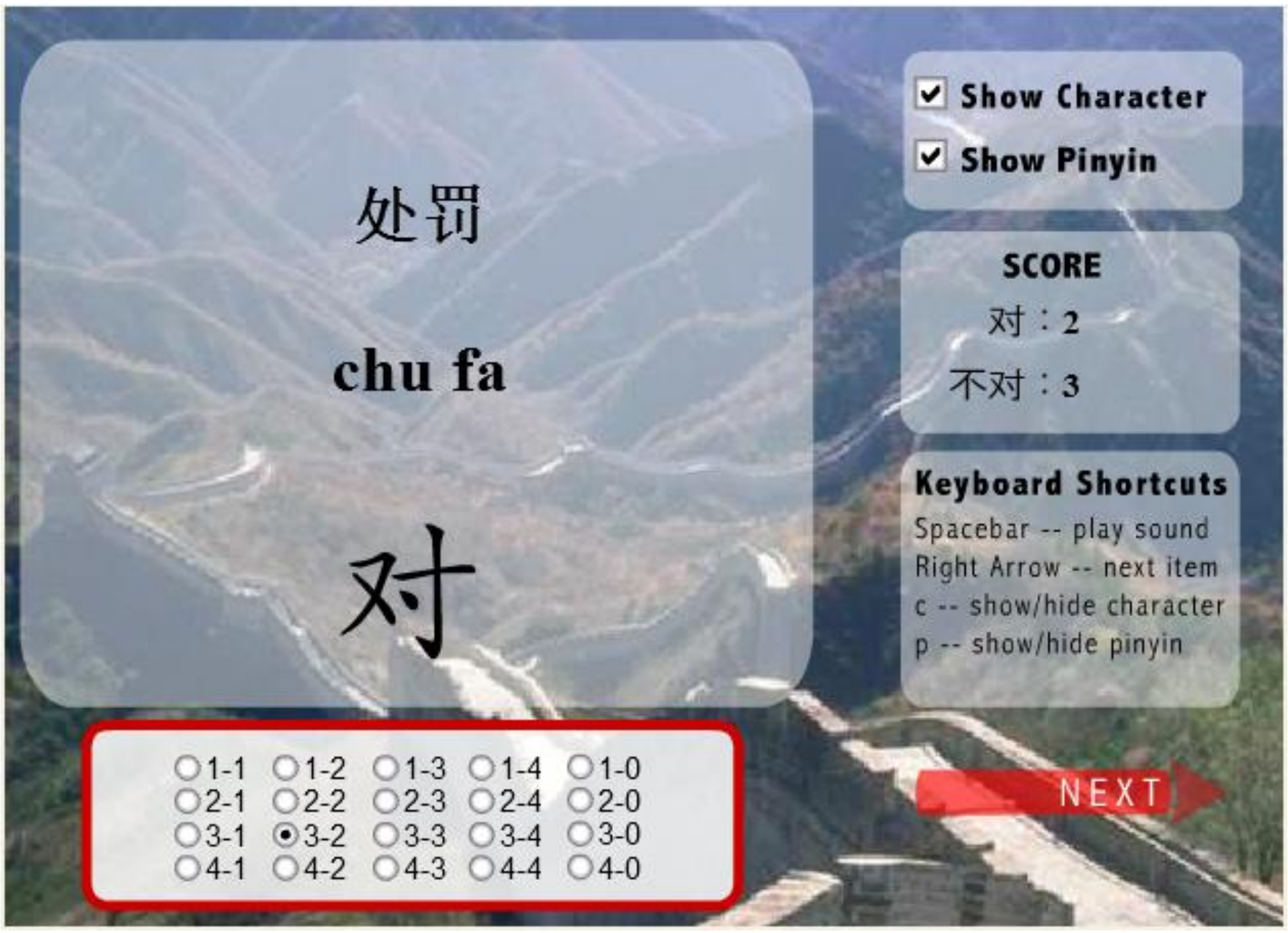

# Appendix C Sample pages of Training Program

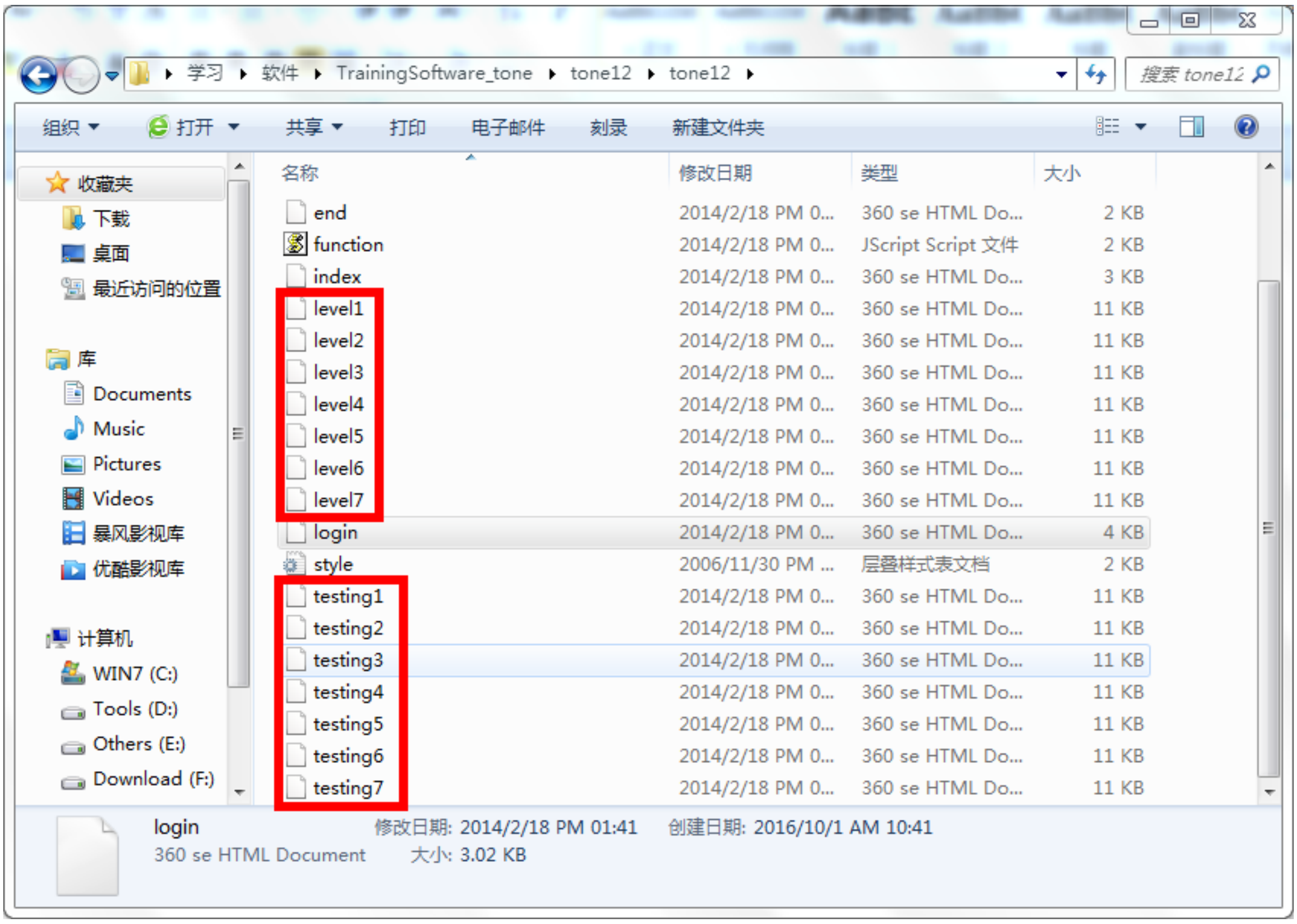

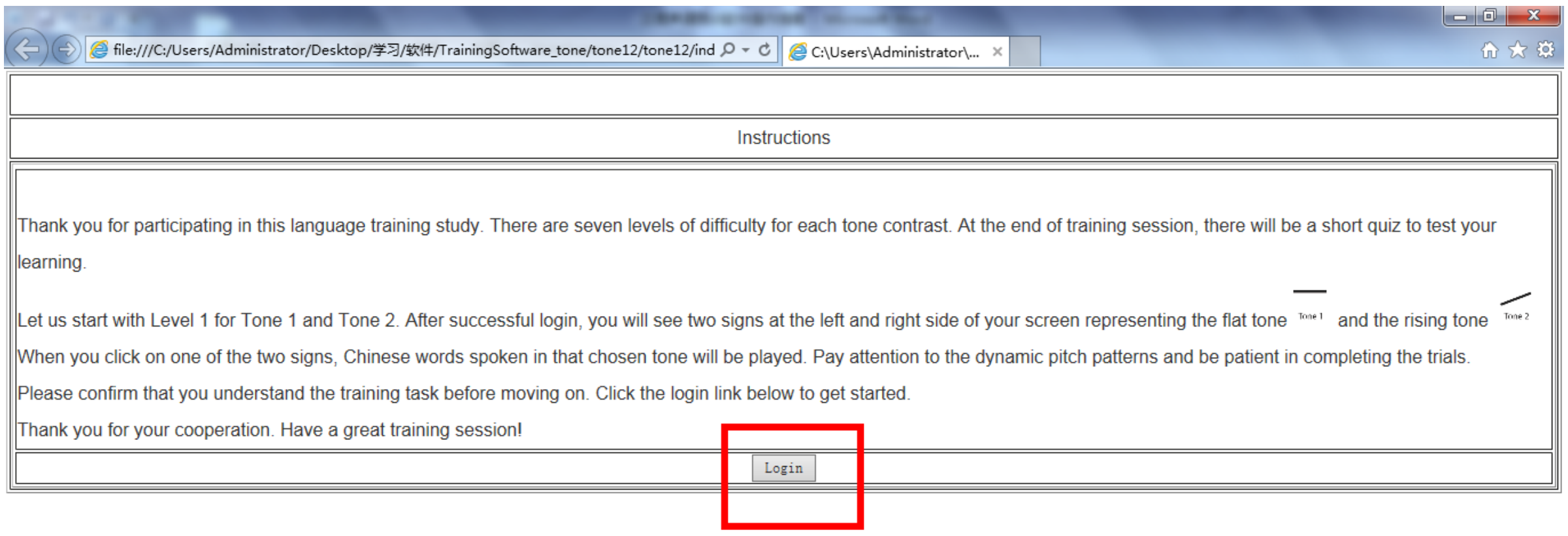

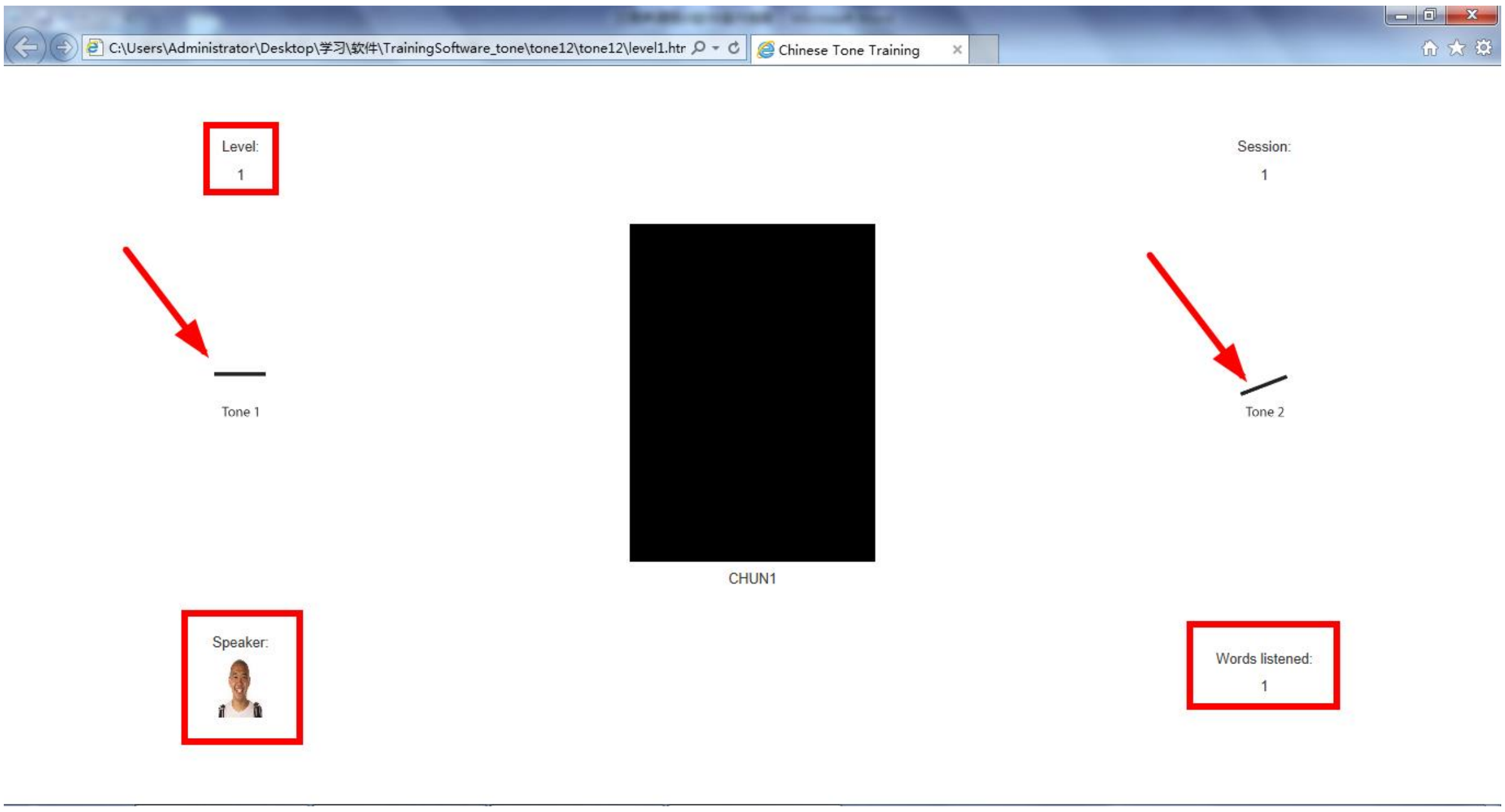
C:\Users\Administrator\Desktop\学习\软件\TrainingSoftware_tone\tone12\tone12\level1.htr
Chinese Tone Training
Level:
1
Session:
1
Tone 1
Tone 2
CHUN1
Speaker:
Words listened:
1

# Appendix D: Survey of Second Language Learning Material

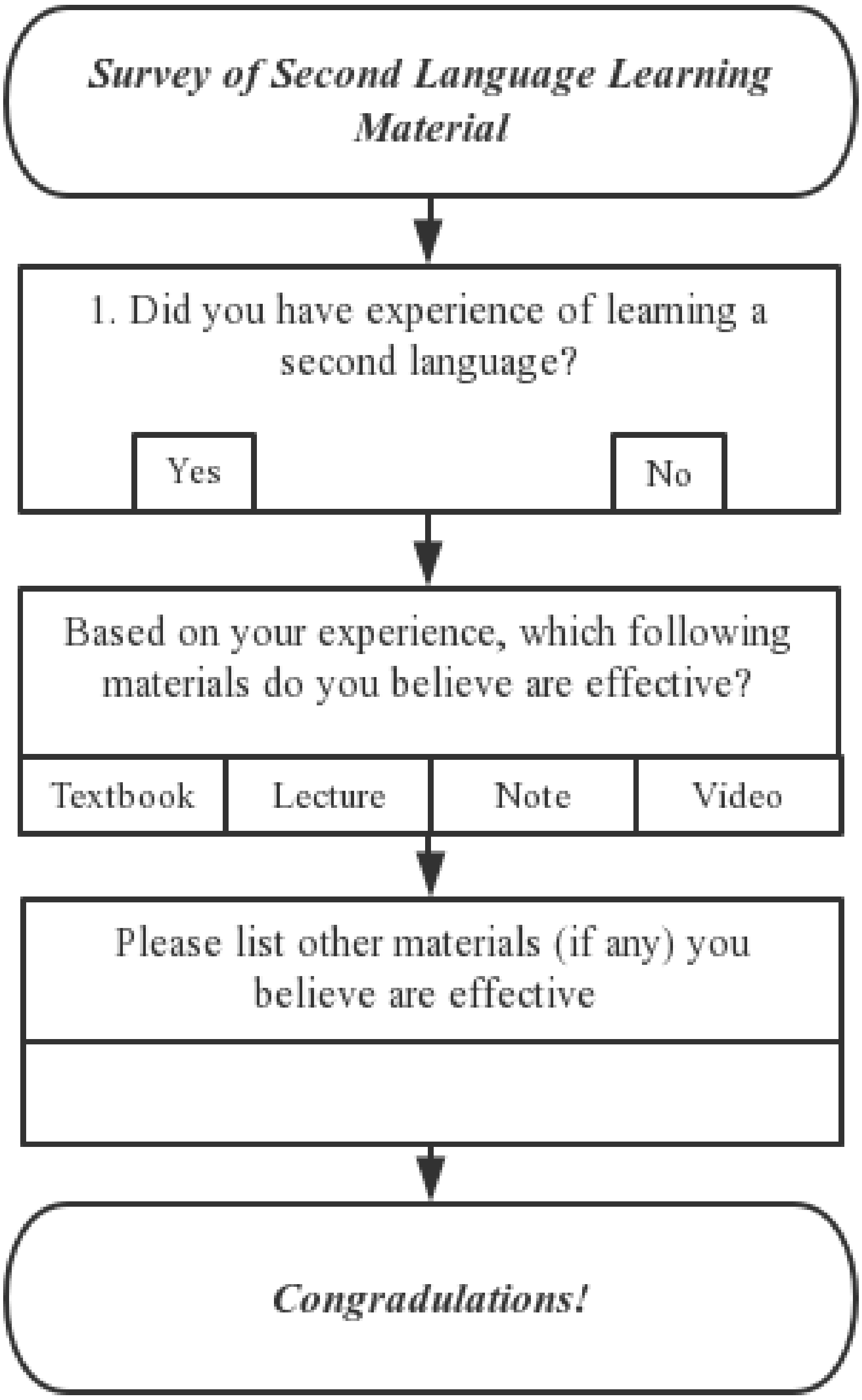

# Appendix E Sample pages of Traditional Training Materials Textbooks

Knowing these travel **cí** will make your holiday twice as enjoyable and at least three times as easy. Review these **cí** by doing the crossword puzzle below. Drill yourself on this Step by selecting other destinations and ask your own **wèntí** about **huǒchē** *(hwoh-chuh)*, **gōnggòngqìchē** *(gohng-gohng-chee-chuh)* **huòzhě** **fēijī** *(fay-jee)* airplanes that go there. Select more **xīn cí** from your **cídiǎn** *(tsih-dee-ahn)* dictionary and practice asking **wèntí** beginning with **nǎr**, **shénme shíhou** and **duōshao qián**.

**ACROSS**
4. sleeping car
6. to book, reserve
7. to arrive, to
8. where
10. international
11. language
12. exit
14. pedicab
15. airport
19. train
20. main station
22. traveler, passenger
25. luggage, baggage
27. time schedule
28. very
29. counter

**DOWN**
1. documents
2. to pay
3. ticket
5. dining car
6. money-exchange office
7. map
9. entrance
12. compartment
13. emergency gate
14. lost-and-found office
16. passport
17. to know
18. airplane
21. to travel
23. fast
24. reclining car
26. boat

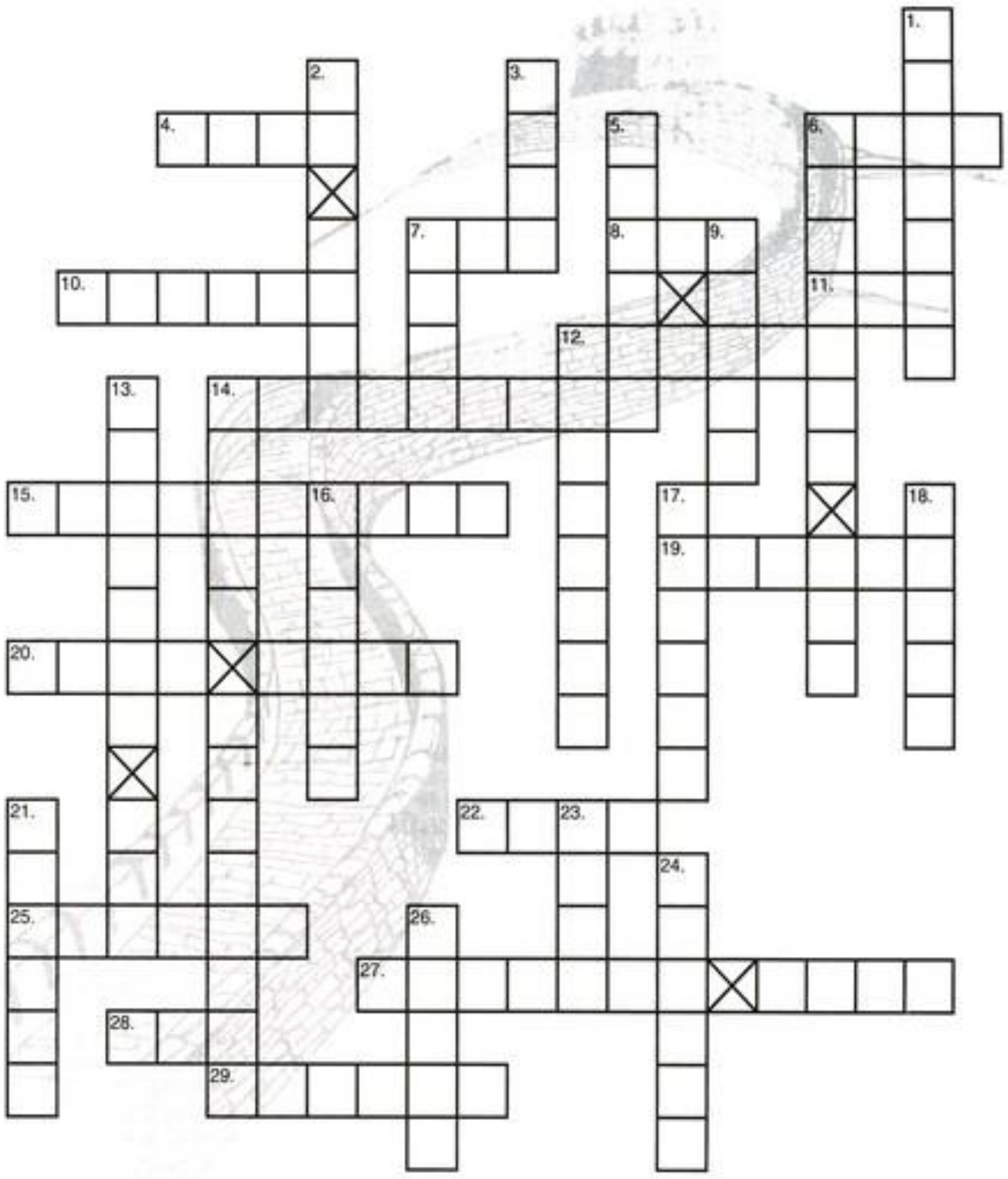

The Great Wall, or as it is known in Chinese — **"Chángchéng,"** was originally constructed during the Qin Dynasty (221-207 B.C.). Over time additional walls were linked to it. The estimates of its exact length range from 1,560 miles to over 3,000 miles.

- ❒ **hǎohàn** *(how-hahn)* .................... wise man, hero ____________
- ❒ **hǎotīng** *(how-teeng)* .................... pleasant to the ear ____________
- ❒ **hǎoxiào** *(how-ssee-ow)* .................... funny ____________
- ❒ **hǎokàn** *(how-kahn)* .................... good-looking ____________
- ❒ **hǎoyùn** *(how-yoon)* .................... good fortune ____________

好 *hǎo*

83

(ssee-ahn-zi) (nee)
**Xiànzài**, it is your turn to practice what **nǐ** have learned. Fill in the following blanks with the correct form of the (dwong-tsih) **dòngcí** (verb). Each time **nǐ** write out the sentence, be sure to say it aloud.

(lie)
**lái**
to come

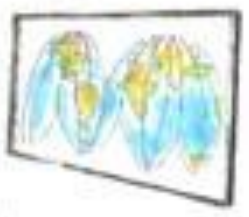

(tswong) (may-gwoh)
Wǒ cóng Měiguó (from, America) ______________ .
(duh-gwoh)
Nǐ cóng Déguó (Germany) ______________ .
(fah-gwoh)
Tā cóng Fǎguó (France) lái/ ______________ .
(yeeng-gwoh)
Wǒmen cóng Yīngguó (England) ______________ .
Tāmen cóng Zhōngguó ______________ .

(ssee-yoo-eh-ssee)
**xuéxi**
to learn

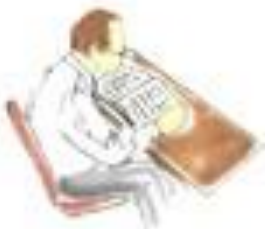

(jwong-wuhn)
Wǒ ______________ Zhōngwén.
(yeeng-wuhn)
Nǐ ______________ Yīngwén (English).
Tā ______________ Zhōngwén.
Wǒmen ______________ Yīngwén.
Tāmen ______________ Zhōngwén.

(choo)
**qù**
to go to

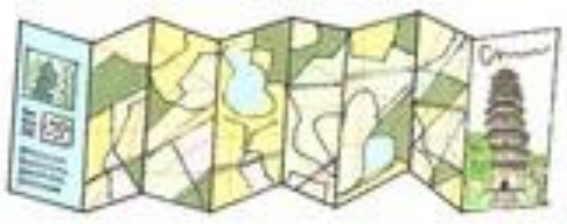

(duh-gwoh)
Wǒ ______________ Déguó (Germany).
(fah-gwoh)
Nǐ ______________ Fǎguó (France).
(yee-dah-lee)
Tā qù/ ______________ Yìdàlì (Italy).
(huh-lahn)
Wǒmen ______________ Hélán (Netherlands).
Tāmen ______________ Zhōngguó.

(yoh)
**yǒu**
to have

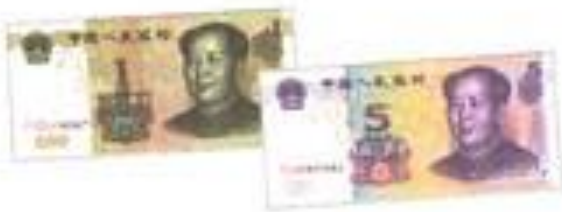

(woo)
Wǒ ______________ wǔ yuán (five yuan).
(lee-oo)
Nǐ ______________ liù yuán (six).
Tā ______________ bā yuán.
(shr)
Wǒmen ______________ shí yuán (ten).
Tāmen ______________ sān yuán.

(ssee-ahng) (yow)
**xiǎng yào**
would like

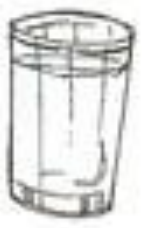

(yee) (bay) (jee-oo)
Wǒ ______________ yì bēi jiǔ (one cup/glass wine).
(chah)
Nǐ ______________ yì bēi chá (cup tea).
(shway)
Tā ______________ yì bēi shuǐ ((M) water).
(joo-zuh-shway)
Wǒmen ______________ yì bēi júzishuǐ ((M) orange juice).
(pee-jee-oo)
Tāmen ______________ yì bēi píjiǔ ((M) beer).

(ssee-oo-yow)
**xūyào**
to need

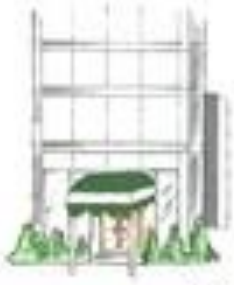

(yee) (jee-ahn) (fahng-jee-ahn)
Wǒ ______________ yì jiān fángjiān (one (M) room).
(yee) (jee-ahn) (fahng-jee-ahn)
Nǐ xūyào/ ______________ yì jiān fángjiān.
Tā ______________ yì jiān fángjiān.
Wǒmen ______________ yì jiān fángjiān.
Tāmen ______________ yì jiān fángjiān.

- ❒ **hǎitān** *(hi-tahn)* ........................ beach
- ❒ **hǎiwài** *(hi-why)* ........................ overseas
- ❒ **hǎiwān** *(hi-wahn)* ........................ bay, gulf
- ❒ **hǎiwèi** *(hi-way)* ........................ seafood
- ❒ **hǎiyáng** *(hi-yahng)* ........................ ocean

海
*hǎi*

43

*(choo-fahng)*
## Chúfáng
kitchen

*(beeng-ssee-ahng)*
**bīngxiāng**
refrigerator

*(loo-zuh)*
**lúzi**
stove

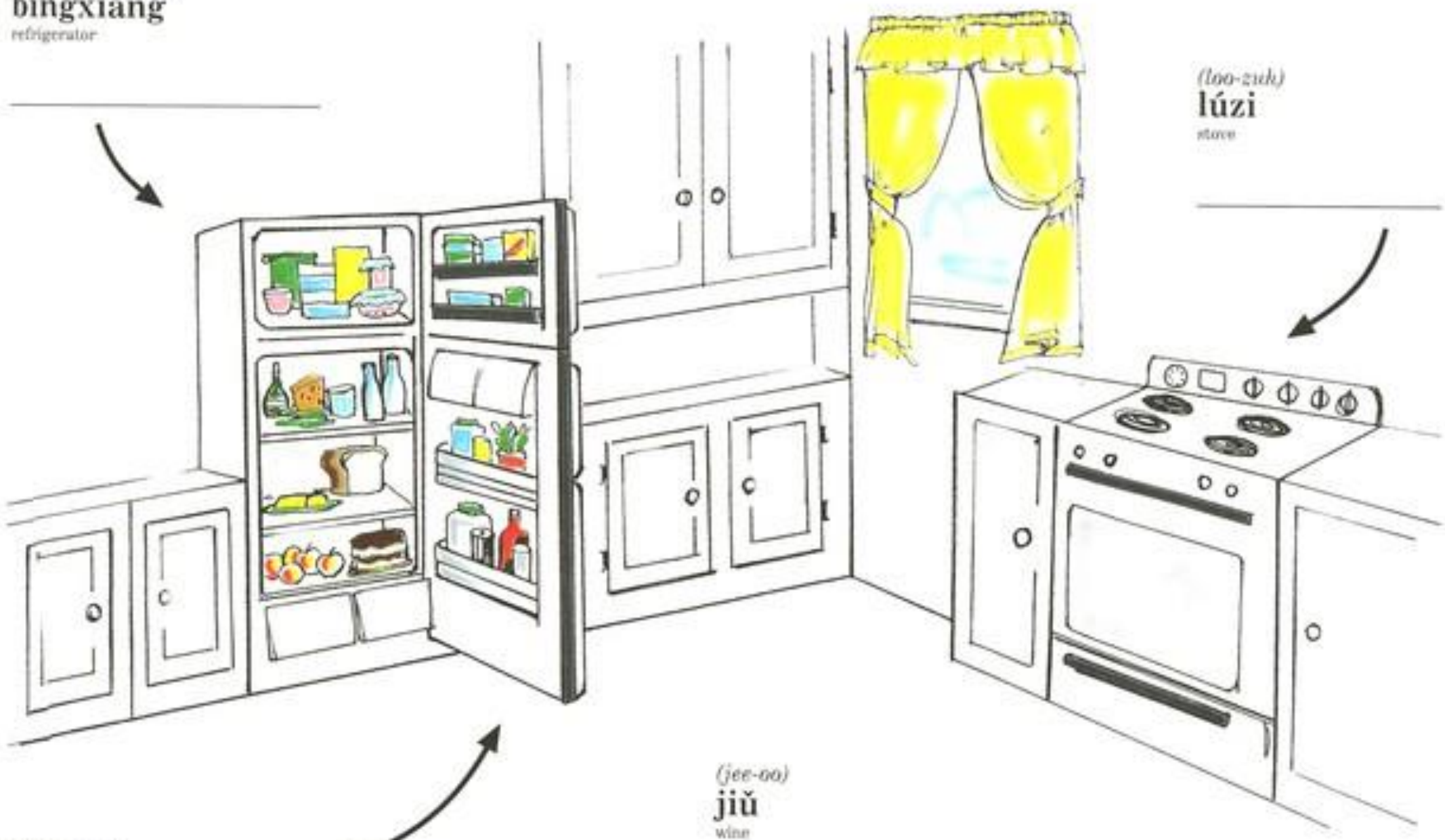

*(jee-oo)*
**jiŭ**
wine

*(nee-oo-yoh)*
**niúyóu**
butter

*(nee-oo-ni)*
**niúnăi**
milk

*(pee-jee-oo)*
**píjiŭ**
beer

Answer these **wèntí** *(wuhn-tee)* (questions) aloud.

**Píjiŭ** *(pee-jee-oo)* (beer) **zài** *(zi)* (is) **năr?** (where) . . . . . . . . . . . . . . . . . . . . . . . . . . . . . . . . . . . . . . . . . **Píjiŭ zài** (is) **bīngxiāng** *(beeng-ssee-ahng)* (refrigerator) **lĭ.** *(lee)* (inside)

**Niúnăi** *(nee-oo-ni)* (milk) **zài** (is) **năr?** (where)

**Jiŭ** *(jee-oo)* (wine) **zài** (is) **năr?** (where)

**Niúyóu** *(nee-oo-yoh)* (butter) **zài** (is) **năr?** (where)

**Kuàngquán** *(kwahng-choo-ahn)* (mineral) **shuĭ** *(shway)* (water) **zài năr?**

**Xiànzài** *(ssee-ahn-zi)* (now) open your **shū** *(shoo)* (book) to the **yè** with the labels and remove the next group of labels and proceed to label all these **dōngxi** *(dwong-ssee)* (things) in your **chúfáng** *(choo-fahng)* (kitchen).

- ❒ **shuĭkù** *(shway-koo)* . . . . . . . . . . . . . . . . . . . . . reservoir
- ❒ **shuĭshŏu** *(shway-shoh)* . . . . . . . . . . . . . . . . . . sailor
- ❒ **shuĭcăi** *(shway-tsi)* . . . . . . . . . . . . . . . . . . . . . . watercolor
- ❒ **shuĭzāi** *(shway-zi)* . . . . . . . . . . . . . . . . . . . . . . flood
- ❒ **shuĭbà** *(shway-bah)* . . . . . . . . . . . . . . . . . . . . . dam

水
*shuĭ*

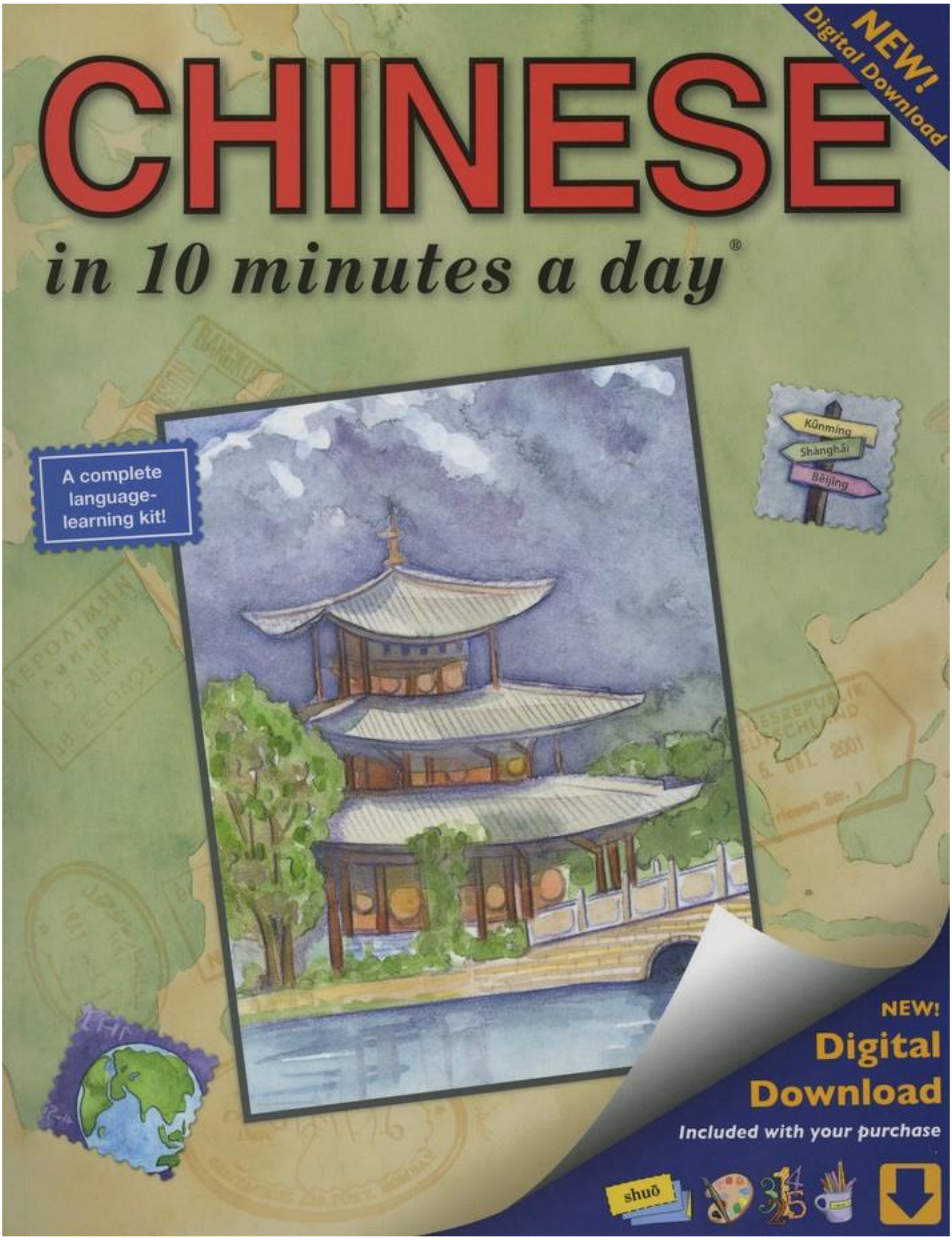
CHINESE
in 10 minutes a day®
NEW! Digital Download
A complete language-learning kit!
Kūnmíng
Shànghǎi
Běijīng
NEW!
Digital
Download
Included with your purchase
shuō

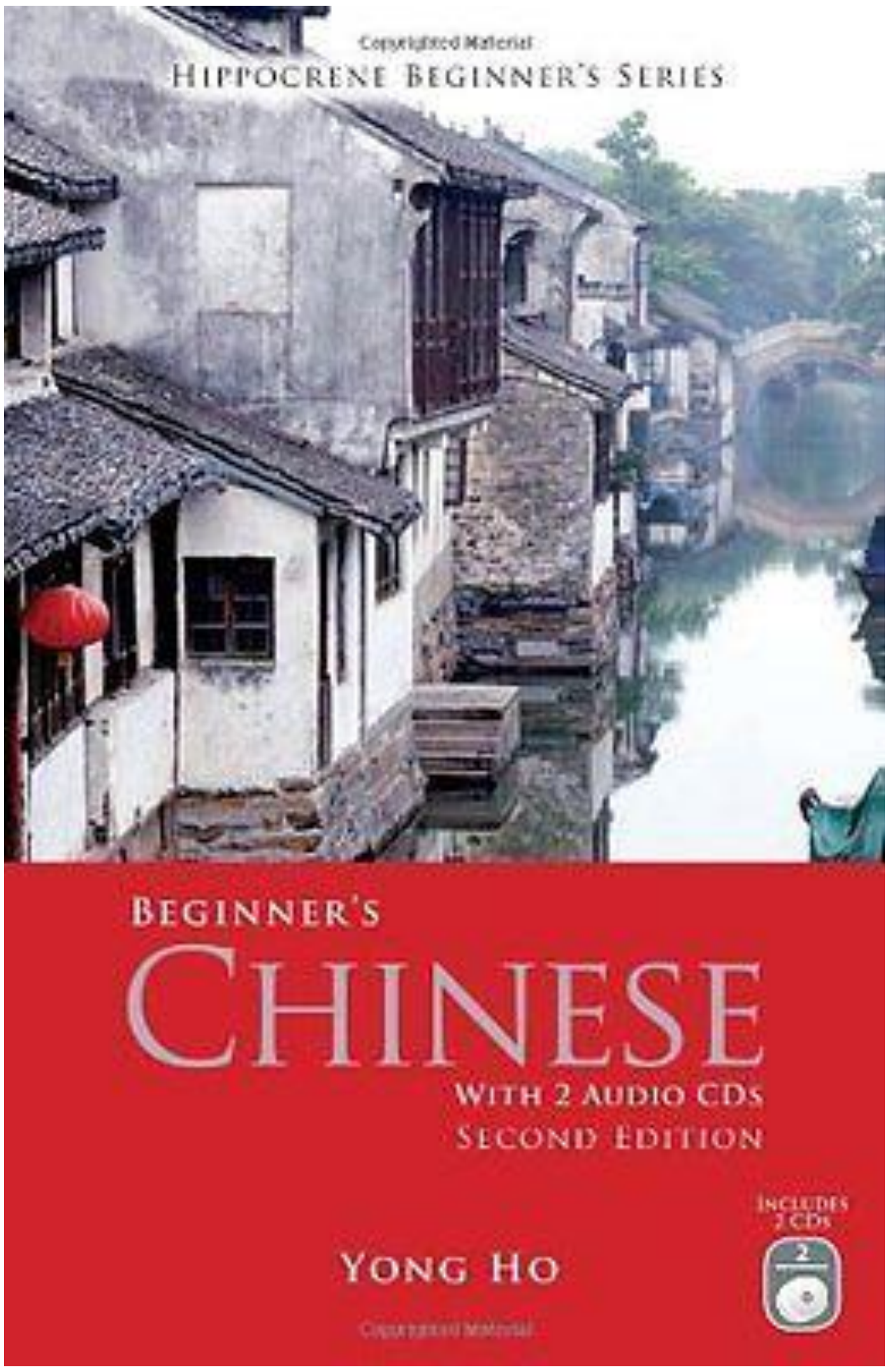

# Appendix F Sample pages of Traditional Training Materials Lectures and Videos

https://www.youtube.com/watch?v=3wV8B4bx1lM)

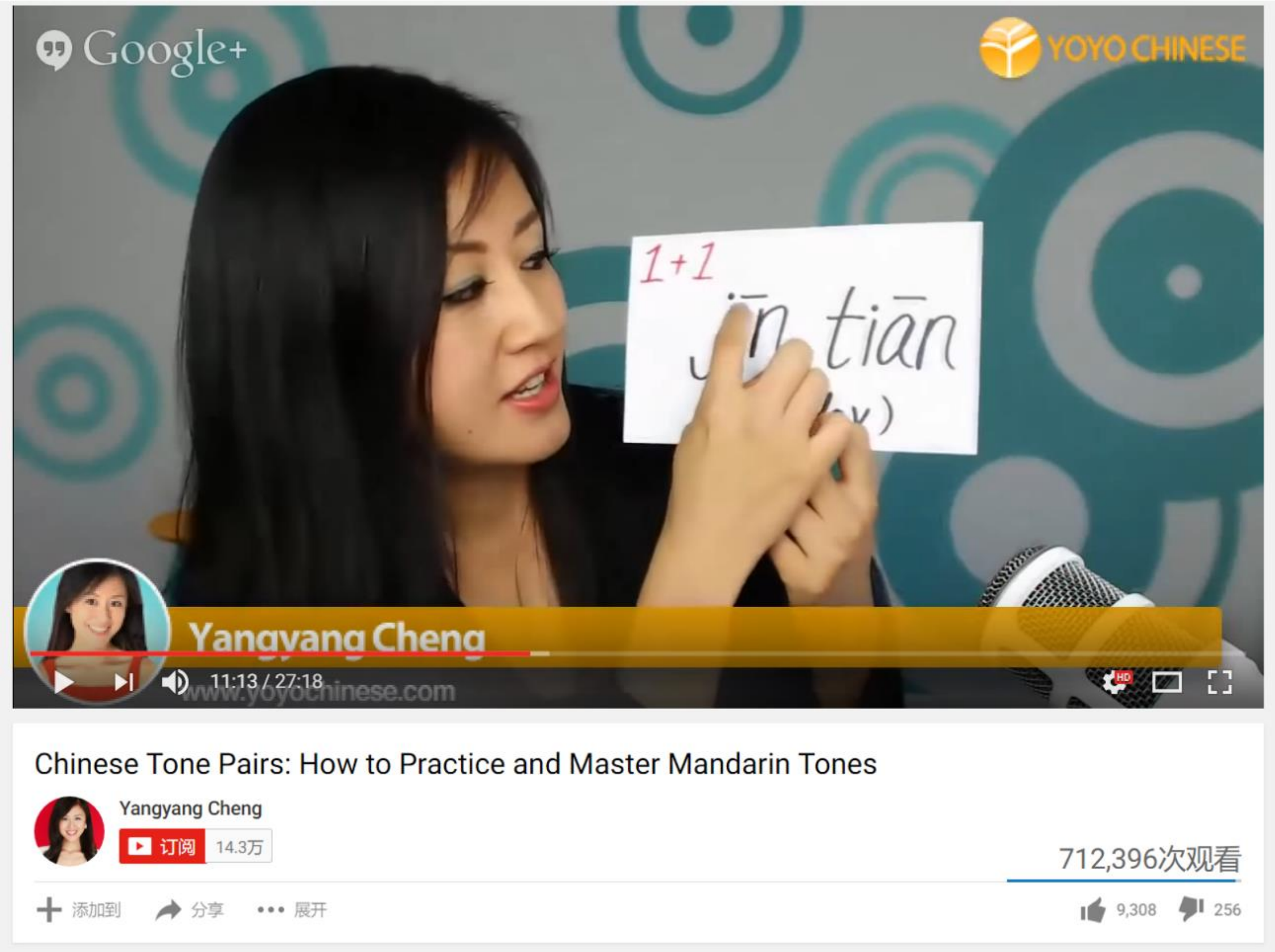

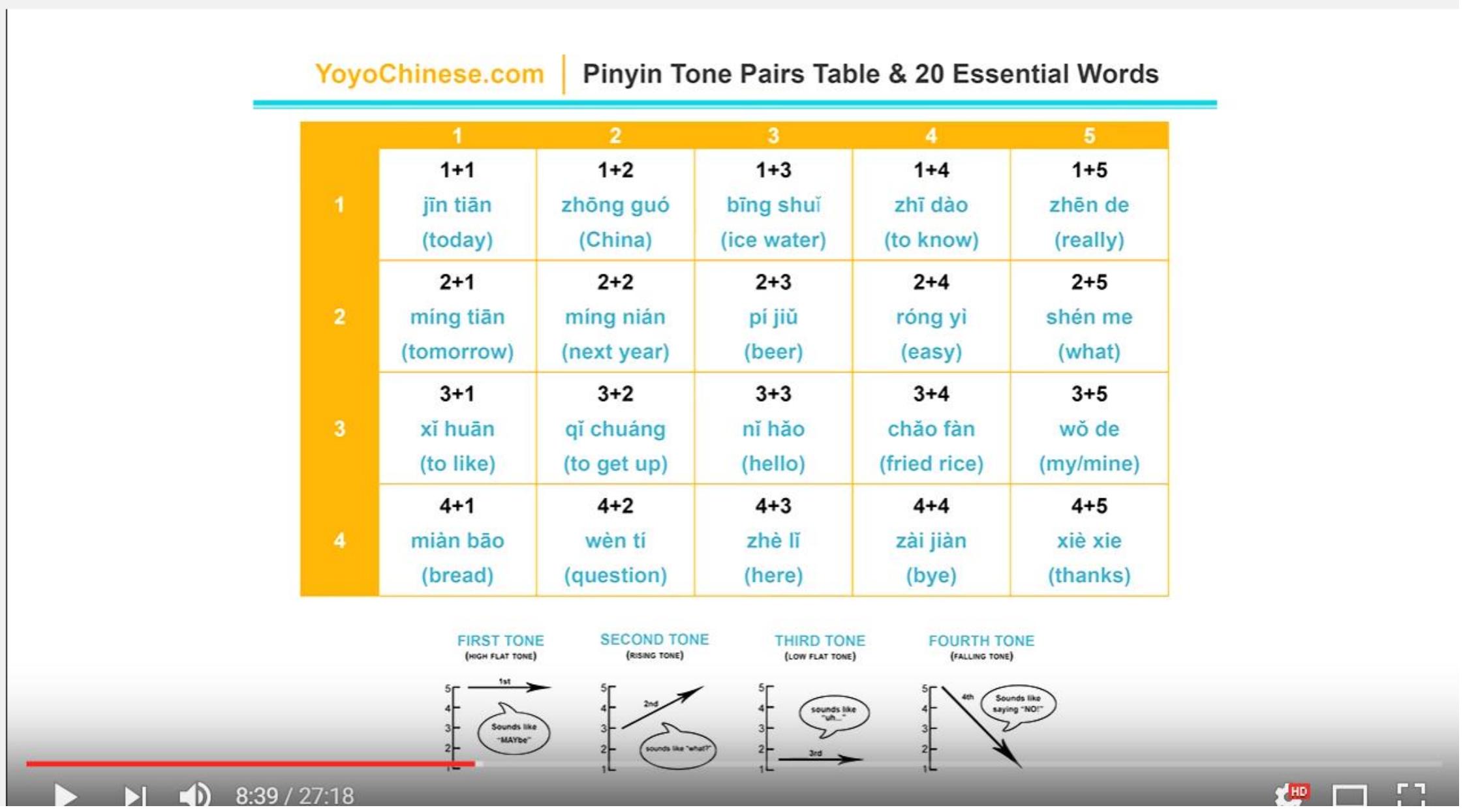

| | 1 | 2 | 3 | 4 | 5 |
|---|---|---|---|---|---|
| 1 | 1+1<br>jīn tiān<br>(today) | 1+2<br>zhōng guó<br>(China) | 1+3<br>bīng shuǐ<br>(ice water) | 1+4<br>zhī dào<br>(to know) | 1+5<br>zhēn de<br>(really) |
| 2 | 2+1<br>míng tiān<br>(tomorrow) | 2+2<br>míng nián<br>(next year) | 2+3<br>pí jiǔ<br>(beer) | 2+4<br>róng yì<br>(easy) | 2+5<br>shén me<br>(what) |
| 3 | 3+1<br>xǐ huān<br>(to like) | 3+2<br>qǐ chuáng<br>(to get up) | 3+3<br>nǐ hǎo<br>(hello) | 3+4<br>chǎo fàn<br>(fried rice) | 3+5<br>wǒ de<br>(my/mine) |
| 4 | 4+1<br>miàn bāo<br>(bread) | 4+2<br>wèn tí<br>(question) | 4+3<br>zhè lǐ<br>(here) | 4+4<br>zài jiàn<br>(bye) | 4+5<br>xiè xie<br>(thanks) |

https://www.youtube.com/watch?v=5HjfI0n7JIM

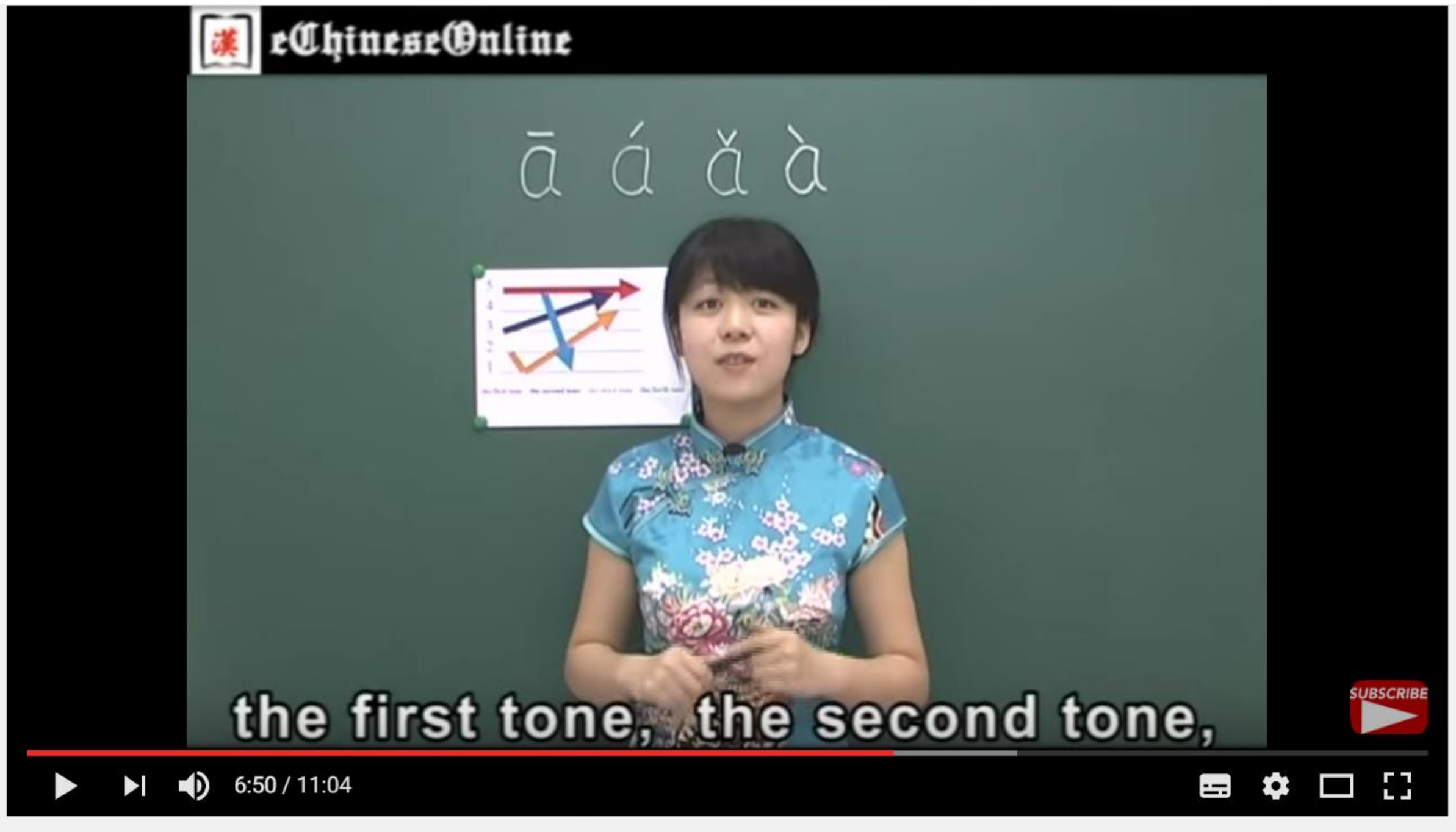

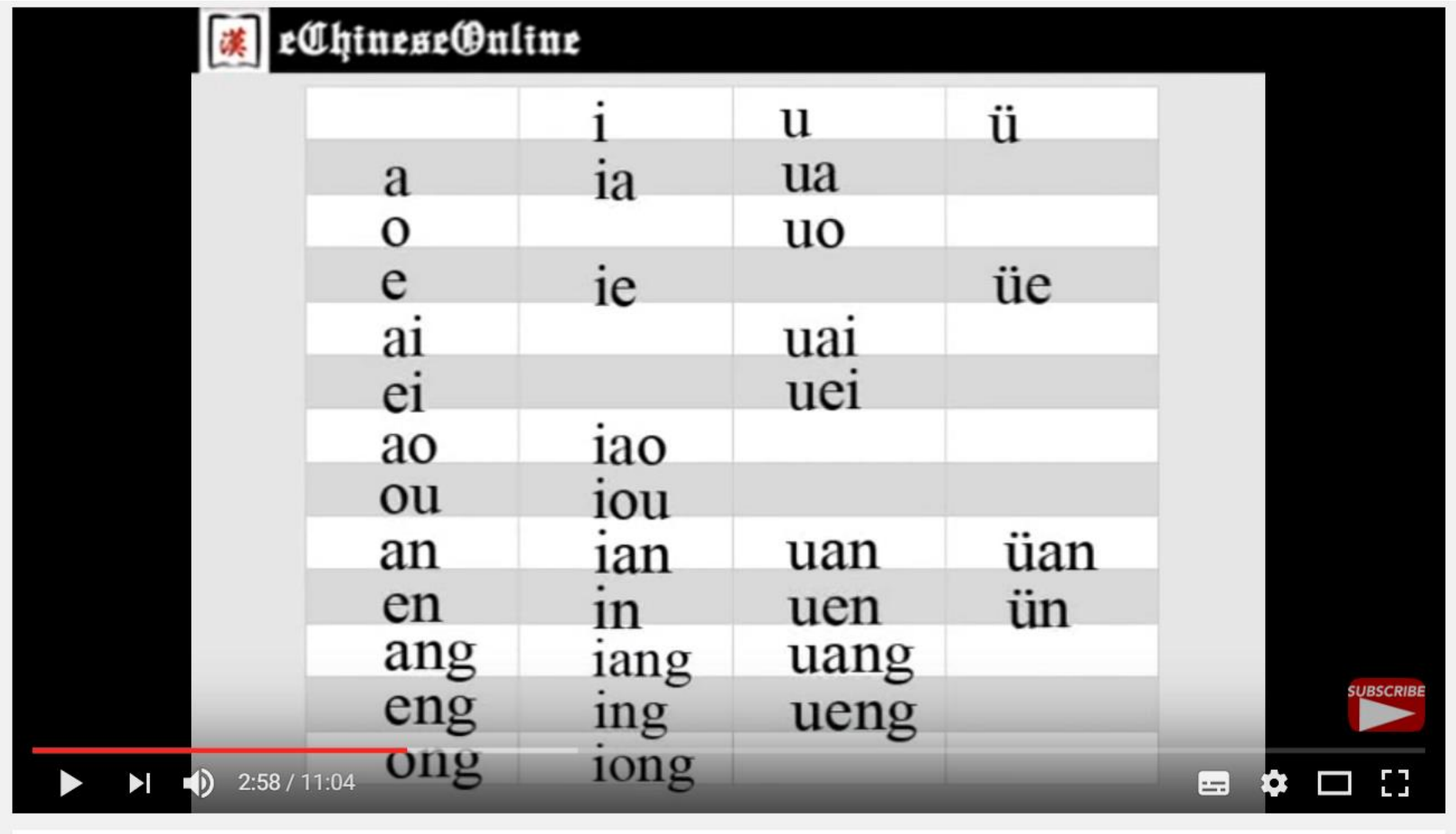

https://www.youtube.com/watch?v=10p2AHD9hmE

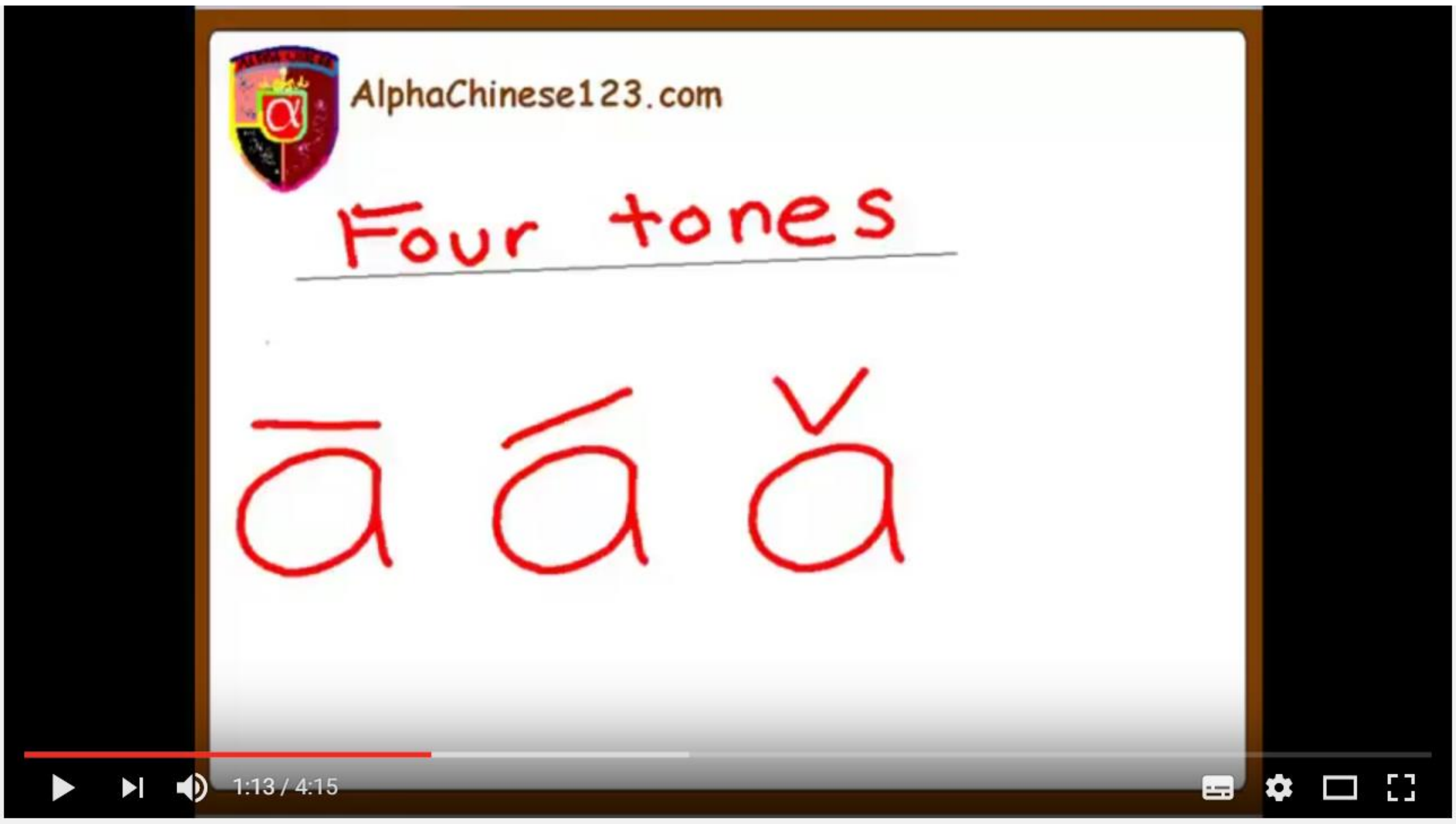

https://www.youtube.com/watch?v=1XQ7MwBmtKQ

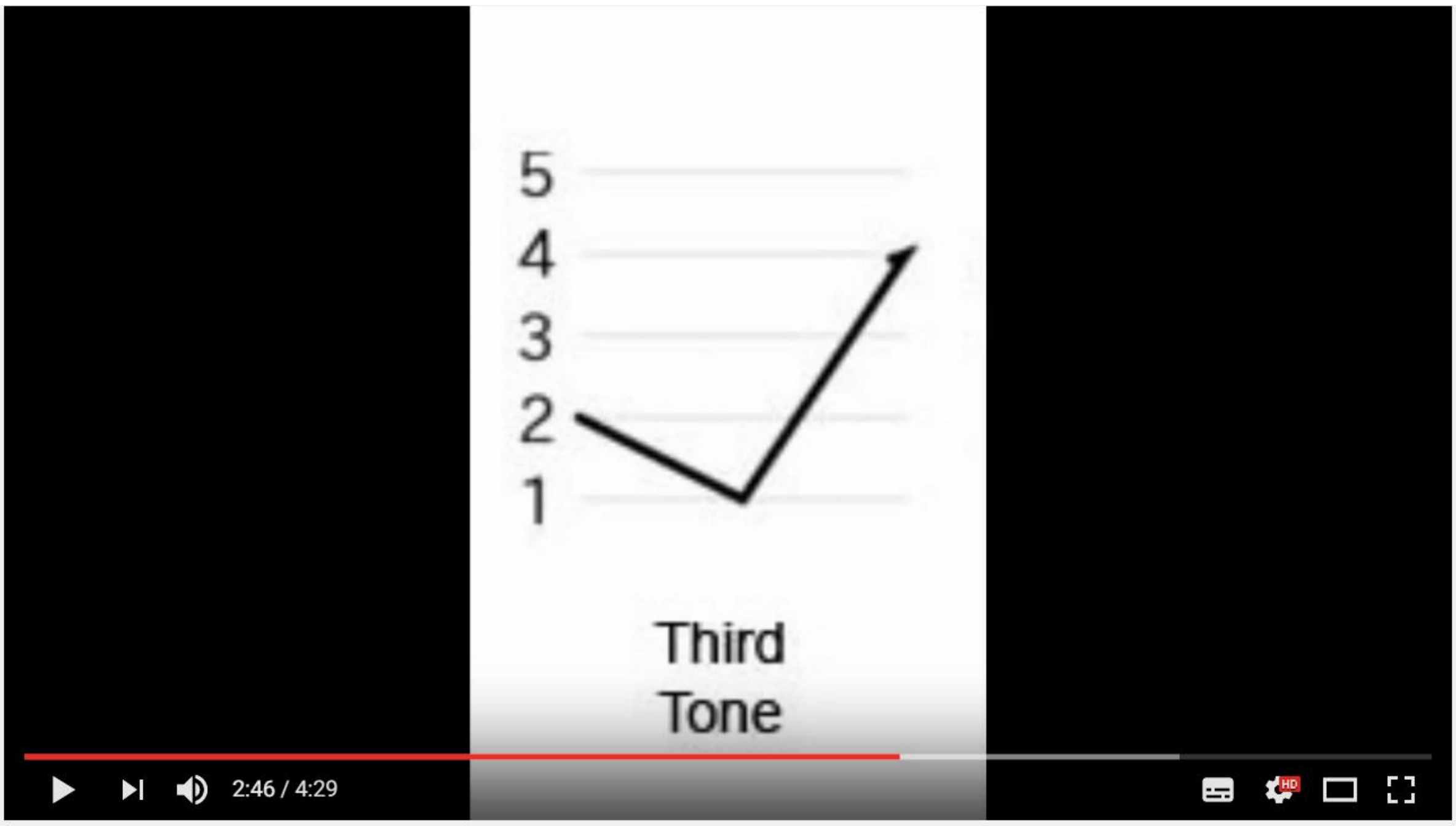

https://www.youtube.com/watch?v=5-_P_H9gMmo

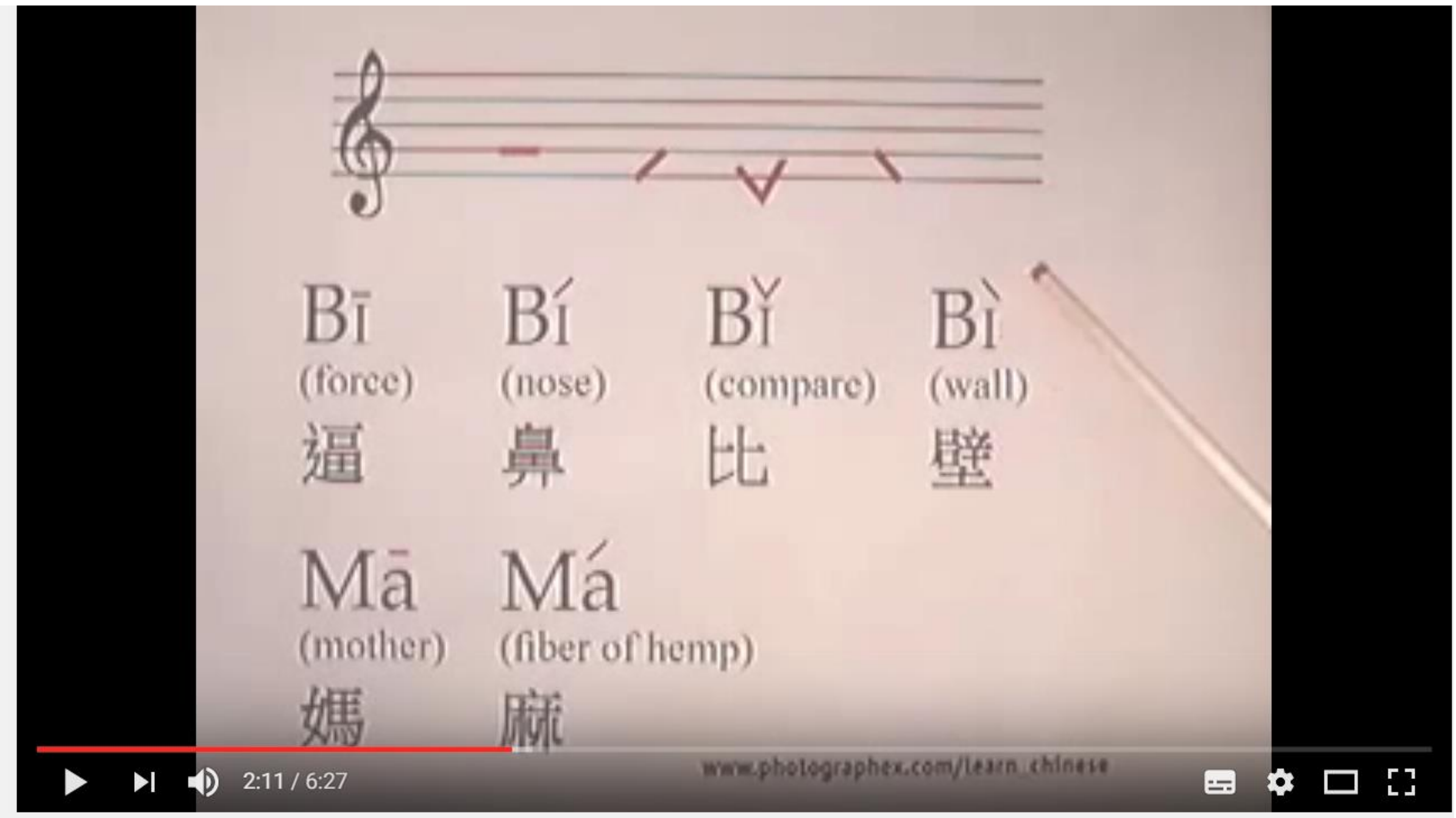

The Four Tones of Mandarin(1st edition, 2006)